\documentclass[acmtog]{acmart}

\usepackage{booktabs} 
\usepackage{subfigure}
\usepackage[inkscapelatex=false]{svg}

\usepackage{lipsum}
\usepackage{amsmath}
\usepackage{scalerel}
\usepackage{bbm}
\usepackage{acro}
\usepackage{bm}
\usepackage{booktabs}
\usepackage{adjustbox}
\usepackage{multirow}
\usepackage{float}
\usepackage{transparent}
\usepackage{subcaption}
\usepackage{tabularx}
\usepackage{makecell}
\usepackage{colortbl}
\usepackage{diagbox}  
\usepackage{algpseudocode}
\usepackage{balance}

\usepackage{glossaries}
\usepackage[table, dvipsnames]{xcolor}
\usepackage{xspace} 

\usepackage[ruled]{algorithm2e} 

\SetAlFnt{\small}
\SetAlCapFnt{\small}
\SetAlCapNameFnt{\small}
\SetAlCapHSkip{0pt}

\definecolor{bestcolor}{HTML}{C0E2CA}
\definecolor{sbestcolor}{HTML}{E2EDB9}

\newcommand{\bmu}{\pmb{\mu}}

\newcommand{\bp}{\mathbf{p}}
\newcommand{\bo}{\mathbf{o}}

\newcommand{\bSigma}{\pmb{\Sigma}}

\newcommand{\our}{\textit{MrHash}}

\def\secref#1{Sec.~\ref{#1}}
\def\figref#1{Fig.~\ref{#1}}
\def\tabref#1{Tab.~\ref{#1}}
\def\eqref#1{Eq.~(\ref{#1})}
\def\algref#1{Alg.~\ref{#1}}

\def\ie{{i.e.}}

\def\lidar{LiDAR}

\def\rgbd{RGB-D}
\def\n3mapping{N$^3$-Mapping}
\DeclareAcronym{tsdf}{
    short = TSDF,
    long = Truncated Signed Distance Function
}
\DeclareAcronym{sdf}{
    short = SDF,
    long = Signed Distance Function
}
\DeclareAcronym{nvs}{
    short = NVS,
    long = Novel View Synthesis
}
\DeclareAcronym{nc}{
    short = NC,
    long = Newer College Dataset
}
\DeclareAcronym{fps}{
    short = FPS,
    long = Frame Per Seconds
}
\DeclareAcronym{psnr}{
    short=PSNR, 
    long=Peak Signal-to-Noise Ratio
}
\DeclareAcronym{ssim}{
    short=SSIM, 
    long=Structure Similarity Index Measure
}
\DeclareAcronym{lpips}{
    short=LPIPS, 
    long=Learned Perceptual Image Patch Similarity
}

\colorlet{colorFst}{Green!25}       %
\colorlet{colorSnd}{SpringGreen!45} %
\colorlet{colorTrd}{Yellow!30}      %

\colorlet{colorSep}{blue!5}         %

\newcommand{\fs}{\cellcolor{colorFst}\bf}   %
\newcommand{\nd}{\cellcolor{colorSnd}}      %

\newcommand{\tablefinalreconstruction}{
\begin{table*}[htbp]
  \centering
  \footnotesize
  \setlength{\tabcolsep}{4pt}
  \renewcommand{\arraystretch}{1.4}
    \begin{tabular}{llccccccc!{\vrule width 0.25pt}@{\hskip 4pt}ccc!{\vrule width 0.5pt}@{\hskip 6pt}ccc}

    \toprule
    & & \makecell{\tt{0000}} & \makecell{\tt{0010}} & \makecell{\tt{0059}} & \makecell{\tt{0106}} & \makecell{\tt{0109}} & \makecell{\tt{0181}} & \makecell{\tt{0207}} & \makecell{\tt{quad}} & \makecell{\tt{math}} & \makecell{\tt{cloi}} & \makecell{Avg\\{\tiny RGB-D}} & \makecell{Avg\\{\tiny LiDAR}} & Avg \\
    \midrule
    \rowcolor{colorSep}
    &&\multicolumn{7}{l}{\textit{\textbf{RGB-D}}}&\multicolumn{3}{l}{\textit{\textbf{LiDAR}}}& & & \\

    \multirow{4}{*}{\rotatebox[origin=c]{90}{PIN-SLAM}}
    & Acc.[cm]$\downarrow$ & 2.680 & 4.069 & 5.789 & 6.540 & 2.802 & 4.161 & 4.004 & 9.942 & 10.271 & 9.082 & 4.292 & 9.765 & 5.934 \\
    & Comp.[cm]$\downarrow$ &   1.078 & \nd 1.060 & 1.464 & 1.732 & \nd 0.862 & \nd 1.648 & \nd 1.085 &   13.764 &   13.960 &  14.584 &   1.276 & 14.103 & 5.124 \\
    & C-L1$\downarrow$ & 1.879 & 2.565 & 3.626 & 4.136 & 1.832 & 2.905 & 2.544 &   11.853 & 12.116 &   11.833 & 2.784 & 11.934 & 5.529 \\
    & F-score[\%]$\uparrow$ & 97.966 & 94.486 & 87.775 & 83.363 & 96.209 &   93.494 & 93.496 &   83.724 & 83.596 &   82.431 & 92.398 & 83.250 & 89.654 \\

    \midrule
    \multirow{4}{*}{\rotatebox[origin=c]{90}{VDBFusion}}
    & Acc.[cm]$\downarrow$ &   2.217 &   2.473 & 4.615 & 5.472 & \nd 1.603 &   4.090 & 2.874 & \fs 6.511 & \nd 8.669 & \fs 6.511 &   3.336 & \fs 7.230 & \nd 4.504 \\
    & Comp.[cm]$\downarrow$ & \fs 0.925 & \fs 0.707 & \fs 1.122 & \fs 0.885 & \fs 0.516 & \fs 0.928 & \fs 0.779 & \fs 11.645 & \fs 12.684 & \fs 12.657 & \fs 0.837 & \fs 12.329 & \fs 4.285 \\
    & C-L1$\downarrow$ & \nd 1.571 & 1.590 & 2.869 & 3.179 & \fs 1.059 &   2.509 & 1.826 & \fs 9.078 & \fs 10.677 & \fs 9.584 & \nd 2.086 & \fs 9.780 & \nd 4.394 \\
    & F-score[\%]$\uparrow$ & 97.052 &   95.687 & 88.896 & 85.161 &   97.659 & 89.723 &   94.692 & \fs 89.532 & \fs 87.157 & \fs 87.963 & 92.696 & \fs 88.217 & \nd 91.352 \\

    \midrule
    \multirow{4}{*}{\rotatebox[origin=c]{90}{Voxblox}}
    & Acc.[cm]$\downarrow$ & 2.120 & 2.718 &   4.156 &   4.137 & 2.583 & 4.246 &   2.678 &   9.824 & \fs 8.619 & 9.284 & \nd 3.234 & 9.242 &   5.037 \\
    & Comp.[cm]$\downarrow$ & 1.759 & 6.168 & 2.145 & 2.304 & 9.045 & 2.638 & 3.817 & 17.083 & 14.756 & 26.207 & 3.982 & 19.349 & 8.592 \\
    & C-L1$\downarrow$ & 1.939 & 4.443 & 3.150 & 3.220 & 5.814 & 3.442 & 3.247 & 13.453 & 11.688 & 17.746 & 3.608 & 14.296 & 6.814 \\
    & F-score[\%]$\uparrow$ & 96.076 & 89.194 &   89.313 &   88.593 & 84.653 & 91.204 & 91.329 & 79.400 &   83.699 & 63.518 & 90.052 & 75.539 & 85.698 \\

    \midrule
    \multirow{4}{*}{\rotatebox[origin=c]{90}{Supereight2$\dagger$}}
    & Acc.[cm]$\downarrow$ & 2.151 & 2.364 &  5.354 & 4.781 &  2.064 & 4.221 & 3.151 & \textcolor{red}{fail} & \textcolor{red}{fail} & \textcolor{red}{fail} & 3.441 & - &  - \\
    & Comp.[cm]$\downarrow$ & 1.457 & 1.294 & 2.002 & 1.484 & 1.183 & 1.983 & 1.470 & \textcolor{red}{fail} & \textcolor{red}{fail} & \textcolor{red}{fail} & 1.553 & - & - \\
    & C-L1$\downarrow$ & 1.804 & 1.829 & 3.678 & 3.132 & 1.623 & 3.102 & 2.310 & \textcolor{red}{fail} & \textcolor{red}{fail} & \textcolor{red}{fail} & 2.497 & - & - \\
    & F-score[\%]$\uparrow$ & 97.478 & 97.399 &  87.040 & 88.131 & 97.073 & 91.905 & 95.252 & \textcolor{red}{fail} & \textcolor{red}{fail} & \textcolor{red}{fail} & 93.468 & - & - \\

    \midrule
    \multirow{4}{*}{\rotatebox[origin=c]{90}{Supereight2$\ddagger$}}
    & Acc.[cm]$\downarrow$ & \nd 1.955 & 2.364 & 5.250 & 4.176 & 2.064 & 3.834 & 3.096 & \textcolor{red}{fail} & \textcolor{red}{fail} & \textcolor{red}{fail} & 3.248 & - &  - \\
    & Comp.[cm]$\downarrow$ & 1.452 & 1.296 & 2.011 & 1.464 & 1.182 & 1.954 & 1.476 & \textcolor{red}{fail} & \textcolor{red}{fail} & \textcolor{red}{fail} & 1.548 & - & - \\
    & C-L1$\downarrow$ & 1.704 & 1.830 & 3.631 & 2.820 & 1.623 & 2.894 & 2.286 & \textcolor{red}{fail} & \textcolor{red}{fail} & \textcolor{red}{fail} & 2.398 & - & - \\
    & F-score[\%]$\uparrow$ & \nd 98.168 & 97.394 & 87.454 & 90.752 & 97.075 & 93.458 & 95.445 & \textcolor{red}{fail} & \textcolor{red}{fail} & \textcolor{red}{fail} & \nd 94.250 & - & - \\

    \midrule
    \multirow{4}{*}{\rotatebox[origin=c]{90}{\n3mapping}}
    & Acc.[cm]$\downarrow$ & \textcolor{red}{fail} & \nd 1.705 & \fs 2.338 & \nd 3.374 &   1.706 & \fs 2.134 & \nd 1.786 & \textcolor{red}{fail} & \textcolor{red}{fail} & \textcolor{red}{fail} & \textcolor{red}{2.174} & - &  - \\
    & Comp.[cm]$\downarrow$ & \textcolor{red}{fail} & 1.335 & 1.664 &   1.390 & 1.133 & 2.616 & 1.222 & \textcolor{red}{fail} & \textcolor{red}{fail} & \textcolor{red}{fail} & \textcolor{red}{1.560} & - & - \\
    & C-L1$\downarrow$ & \textcolor{red}{fail} & \nd 1.520 & \fs 2.001 & \nd 2.382 &   1.420 & \fs 2.375 & \nd 1.504 & \textcolor{red}{fail} & \textcolor{red}{fail} & \textcolor{red}{fail} & \textcolor{red}{1.867} & - & - \\
    & F-score[\%]$\uparrow$ & \textcolor{red}{fail} & \fs 98.811 & \fs 96.398 & \nd 93.671 & \nd 98.076 & \fs 97.034 & \fs 97.293 & \textcolor{red}{fail} & \textcolor{red}{fail} & \textcolor{red}{fail} & \textcolor{red}{96.880} & - & - \\

    \midrule
    \multirow{4}{*}{\rotatebox[origin=c]{90}{Ours}}
    & Acc.[cm]$\downarrow$ & \fs 0.963 & \fs 1.587 & \nd 2.981 & \fs 2.822 & \fs 1.364 & \nd 2.789 & \fs 1.701 & \nd 7.134 &   9.061 & \nd 7.015 & \fs 2.030 & \nd 7.737 & \fs 3.742 \\
    & Comp.[cm]$\downarrow$ & \nd 1.053 &   1.079 & \nd 1.414 & \nd 1.202 &  0.961 & 1.968 &  1.234 & \nd 11.898 & \nd 13.116 &   \nd 14.208 & \nd 1.273 & \nd 13.074 & \nd 4.813 \\
    & C-L1$\downarrow$ & \fs 1.008 & \fs 1.333 & \nd 2.198 & \fs 2.012 & \nd 1.162 & \nd 2.379 & \fs 1.366 & \nd 9.516 & \nd  11.088 & \nd 10.612 & \fs 1.637 & \nd 10.405 & \fs 4.267 \\
    & F-score[\%]$\uparrow$ & \fs 99.674 & \nd 98.408 & \nd 94.124 & \fs 94.200 & \fs 98.079 & \nd 96.183 & \nd 97.246 & \nd 89.503 & \nd 85.811 & \nd 84.815 & \fs 96.845 & \nd  86.710 & \fs 93.804 \\

    \bottomrule
  \end{tabular}
  \caption{\textbf{3D Reconstruction Results}. The \rgbd~data are sequences of ScanNet~\cite{dai2017scannet} and \lidar~data are sequences of NC~\cite{zhang2021multicamera}. All the pipelines are run with ground-truth fixed poses. Voxel size is set to 1 cm for \rgbd~data and 20 cm for \lidar~data, and the F-score is computed with a 10 cm error threshold for the \rgbd~and 20 cm for the \lidar. Best results are highlighted as \colorbox{colorFst}{\textbf{first}} and \colorbox{colorSnd}{second}. Supereigth$\dagger$ represents single resolution grid, Supereight$\ddagger$ represents multi-resolution grid. Both Supereight2$\dagger$ and Supereight2$\ddagger$ failed on all \lidar~sequences. \n3mapping fails on one \rgbd~sequence and all the \lidar~sequences; for completeness, we report the average for \rgbd~despite is calculated only on its working sequences.}
  \label{tab:final_reconstruction}
\end{table*}
}

\newcommand{\tableruntimes}{
\begin{table}[htbp]
  \centering
  \scriptsize
  \setlength{\tabcolsep}{1pt}
  \renewcommand{\arraystretch}{1.4}
  \begin{tabular}{lcc|cc|cc|cc}
    \toprule
    \rowcolor{colorSep}
    \textbf{Method} 
    & \multicolumn{2}{|c|}{\textbf{ScanNet}} 
    & \multicolumn{2}{c|}{\textbf{Replica}} 
    & \multicolumn{2}{c|}{\textbf{Oxford-Spires}} 
    & \multicolumn{2}{c}{\textbf{NC}} \\
    \cmidrule(r){1-1} \cmidrule(lr){2-3} \cmidrule(lr){4-5} \cmidrule(lr){6-7} \cmidrule(lr){8-9}
    & Time [ms] $\downarrow$ & FPS $\uparrow$
    & Time [ms] $\downarrow$ & FPS $\uparrow$
    & Time [ms] $\downarrow$ & FPS $\uparrow$
    & Time [ms] $\downarrow$ & FPS $\uparrow$ \\
    \midrule
    VDBFusion         & 17.45 & 5.71  & 519  & 1.93   & 40.31  & 24.65 & 103.62   & 9.61 \\
    PIN-SLAM          & 67.15 & 14.80 & 68  & 14.7   & 75.99  & 13.15 & 113.68   & 8.79 \\
    \n3mapping        & 184.67 & 5.40  & --  & --   & 255.44  & 3.91  & --   & -- \\
    NKSR              & --  & --    & --  & --   & \nd 16.05 & \nd 62.28 & --   & -- \\
    Voxblox           & 132.06 & 7.57  & 373  & 2.68   & 61.72  & 16.17 & 162.47   & 6.14 \\
    Supereight2$\dagger$ & 73.94 & 13.53  & 100.87  & 9.92   & --  & -- & --   & -- \\
    Supereight2$\ddagger$ & 79.52 & 12.55 & 98.11  & 10.20   & --  & -- & --   & -- \\
    Ours (single)     & \fs 15.11 & \fs 64.30 & \fs 20.45 & \fs 48.9 & \fs 14.27 & \fs 68.86 & \nd 30.18 & \nd 21.6 \\
    Ours (multi)      & 16.79 \nd  & \nd 59.34    & \nd 37.50   & \nd 26    & 16.39    & 61    & \fs 28.17   &  \fs 35.25 \\
    \bottomrule
  \end{tabular}
  \caption{\textbf{Runtime comparison.} Time [ms] refers to the total computation time for one frame of the sequence, while FPS refers to processing speed. Best results are highlighted as \colorbox{colorFst}{\textbf{first}} and \colorbox{colorSnd}{second}. Dash ``-'' indicates missing values due to failure. Supereigth$\dagger$ represents single resolution grid, Supereight$\ddagger$ represents multi-resolution grid. Our runtimes show an improvement from 2-3x to 13x compared to existing SOTA.}
  \label{tab:runtimes}
\end{table}
}

\newcommand{\tablememory}{
\begin{table*}[htbp]
  \centering
  \footnotesize
  \setlength{\tabcolsep}{4pt}
  \renewcommand{\arraystretch}{1.4}
  \begin{tabular}{lccc|ccc|ccc}
    \toprule
    \rowcolor{colorSep}
    \textbf{Method} 
    & \multicolumn{3}{c|}{\textbf{RGB-D}} 
    & \multicolumn{3}{c|}{\textbf{LiDAR}} & \multicolumn{3}{c}{\textbf{Avg.}} \\
    \cmidrule(r){1-1} \cmidrule(lr){2-4} \cmidrule(lr){5-7} \cmidrule(lr){8-10}
    & \# Vertices $\downarrow$ & \#  Faces $\downarrow$ & Memory [MB] $\downarrow$ & \#  Vertices $\downarrow$ & \# Faces $\downarrow$ & Memory [MB] $\downarrow$ & \#  Vertices $\downarrow$ & \# Faces $\downarrow$ & Memory [MB] $\downarrow$ \\
    \midrule
    VDBFusion     &  4576631 & 8575618 & 331.2 &  \fs 321179 &  \fs 587890 & \fs 15.4 & 2448905 & 4581754 & 173.3 \\
    PIN-SLAM      & 9160908 & 16851839 & 658.8 & 896877 & 1626551 & 64.2  & 5028893 & 9239195 & 361.5 \\
    Voxblox       & 5841313 & 7168535 & 373.6 & 828294 & 845420 & 50.7  & 3334804 & 4006978 & 212.15 \\
    Supereight2$\dagger$       & 27171273 & 9057091 & 1163.4 & -- & -- & --  & -- & -- & -- \\
    Supereight2$\ddagger$       & 19247955 & 6415985 & 817.0 & -- & -- & --  & -- & -- & -- \\
    Ours (single) & \nd 4501002 &  \nd 4070646 & \nd 242.2 & 753320 & 1246638 & 22.8  & \nd 2431648 & \nd 2349585 & \nd 132.5 \\
    Ours (multi)  &  \fs 2068369 &  \fs 3752706 & \fs 155.7 & \nd 615204 & \nd 1035968 & \nd 21.60  & \fs 1206970 & \fs 2173218 & \fs 88.65 \\
    \bottomrule
  \end{tabular}
  \caption{\textbf{Average memory usage.} We report the number of vertices and faces as a proxy for memory consumption, as they correlate with mesh complexity and storage requirements. Best values are marked as \colorbox{colorFst}{\textbf{first}} and \colorbox{colorSnd}{second}. Dash ``-'' indicates missing values due to failure. Supereigth$\dagger$ represents resolution grid, Supereight$\ddagger$ represents multi-resolution grid. Our multi-resolution setup improves from 2.0x to 7.5x in the \rgbd~setup and from 2.3x to 2.9x in the \lidar~setup (excluding VDBFusion meshes that are 0.7x compared to ours). On average, the improvement is from 1.95x to 4.07.}
  \label{tab:memory_avg}
\end{table*}
}

\newcommand{\tablegswithtiming}{
\begin{table}[htbp]
  \centering
  \footnotesize
  \setlength{\tabcolsep}{6pt}
  \renewcommand{\arraystretch}{1.4}
  \begin{tabular}{lccccc}
    \toprule
    \rowcolor{colorSep}
    \textbf{Method} & PSNR $\uparrow$ & SSIM $\uparrow$ & LPIPS $\downarrow$ & FPS $\uparrow$ & Time[ms] $\downarrow$ \\
    \midrule
    GSFusion & \nd 34.65 & 0.949 & 0.056 & 14.99 & 66.67 \\
    Ours & 33.90 & 0.949 &  0.057 & \fs 28.05 & \fs 35.65 \\
    Ours (+iterations) & 34.27 & \nd 0.951 & \nd 0.052 & 14.22 & 70.32 \\
    Ours (multi) & \fs 35.73 &	\fs 0.960 &	\fs 0.044 & \nd 23.70 & \nd 42.19  \\
    \bottomrule
  \end{tabular}
  \caption{\textbf{Rendering quality and performance}. Comparison of perceptual metrics (\ac{psnr}, \ac{ssim}, \ac{lpips}) and rendering speed (\ac{fps}) on Replica dataset \cite{straub2019replica}. Best results are highlighted as \colorbox{colorFst}{\textbf{first}} and \colorbox{colorSnd}{second}. Ours(+iterations) means we set more iterations to optimize GS parameters, using the single-resolution grid. The multi-resolution setup improves overall rendering quality in the GS pipeline, with a small computational overhead. Nevertheless, our GPU-based quad-tree implementation achieves better overall performance compared to other GSFusion.}
  \label{tab:render_gs}
\end{table}
}

\newcommand{\tablegswrecon}{
\begin{table}[htbp]
  \centering
  \scriptsize
  \setlength{\tabcolsep}{6pt}
  \renewcommand{\arraystretch}{1.4}
  \begin{tabular}{lccccc}
    \toprule
    \rowcolor{colorSep}
    \textbf{Method} & Mesh [MB] $\downarrow$ & Acc.[cm] $\downarrow$ & Comp.[cm] $\downarrow$ & C-L1 $\downarrow$ & F-Score[\%] $\uparrow$ \\
    \midrule
    GSFusion & \nd 454.025 & \fs 0.608 & \fs 2.777 & \fs 1.692 & \fs 95.192 \\
    Ours & \fs 132.650 & \nd 0.614 & \nd 2.832 & \nd 1.723 & \nd 95.106 \\
    \bottomrule
  \end{tabular}
  \caption{\textbf{Reconstruction quality and performance.} We compare the two approaches on the same sequences of \tabref{tab:render_gs} \ie, Replica dataset \cite{straub2019replica}. Best results are highlighted as \colorbox{colorFst}{\textbf{first}} and \colorbox{colorSnd}{second}. Considering the synthetic nature of the data, the results are almost equal at mm level. Nevertheless, the mesh size of GFusion is almost 3x times bigger than ours.}
  \label{tab:gsfusion_recon}
\end{table}
}

\newcommand{\tablemultimethod}{
\begin{table*}[htbp]
  \centering
  \footnotesize
  \setlength{\tabcolsep}{4pt}
  \renewcommand{\arraystretch}{1.4}
  \begin{tabular}{llccccccc!{\vrule width 0.25pt}@{\hskip 4pt}ccc!{\vrule width 0.5pt}@{\hskip 6pt}ccc}
    \toprule
    & & \makecell{\tt{0000}} & \makecell{\tt{0010}} & \makecell{\tt{0059}} & \makecell{\tt{0106}} & \makecell{\tt{0109}} & \makecell{\tt{0181}} & \makecell{\tt{0207}} & \makecell{\tt{quad}} & \makecell{\tt{math}} & \makecell{\tt{cloi}} & \makecell{Avg\\{\tiny RGB-D}} & \makecell{Avg\\{\tiny LiDAR}} & Avg \\
    \midrule
    \rowcolor{colorSep}
    &&\multicolumn{7}{l}{\textit{\textbf{RGB-D}}}&\multicolumn{3}{l}{\textit{\textbf{LiDAR}}}& & & \\

    \multirow{4}{*}{\rotatebox[origin=c]{90}{Ours}}
    & Acc.[cm]$\downarrow$ &   0.963 &   1.587 &   2.981 &   2.822 &   1.364 &   2.789 &   1.701 &   7.134 &   9.061 &   7.015 &   2.030 &   7.737 &   3.742 \\
    & Comp.[cm]$\downarrow$ & 1.053 &   1.079 &   1.414 &   1.202 &   0.961 &   1.968 & 1.234 &   11.898 &   13.116 &   14.208 &   1.273 &   13.074 &   4.813 \\
    & C-L1$\downarrow$ &   1.008 &   1.333 &   2.198 &   2.012 &   1.162 &   2.379 &   1.366 &   9.516 &   11.088 &   10.612 &   1.637 &   10.405 &   4.267 \\
    & F-score[\%]$\uparrow$ &   99.674 &   98.408 &   94.124 &   94.200 &   98.079 &   96.183 &   97.246 &   89.503 &   85.811 &   84.815 &   96.845 &   86.710 &   93.804 \\

    \midrule
    \multirow{4}{*}{\rotatebox[origin=c]{90}{Ours (Multi)}}
    & Acc.[cm]$\downarrow$ & 1.207 & 1.988 & 3.688 & 3.671 & 1.757 & 3.424 & 2.047 & 10.763 & 12.314 & 10.326 & 2.540 & 11.135 & 5.119 \\
    & Comp.[cm]$\downarrow$ &   1.036 & 1.297 & 1.433 & 1.372 & 1.183 & 2.006 &   1.088 & 18.810 & 16.977 & 22.107 & 1.345 & 19.298 & 6.731 \\
    & C-L1$\downarrow$ & 1.121 & 1.642 & 2.560 & 2.522 & 1.470 & 2.715 & 1.567 & 14.786 & 14.646 & 16.217 & 1.943 & 15.216 & 5.925 \\
    & F-score[\%]$\uparrow$ & 99.503 & 98.127 & 92.575 & 91.716 & 98.029 & 94.902 & 97.094 & 81.002 & 77.864 & 76.170 & 95.992 & 78.346 & 90.698 \\

    \bottomrule
  \end{tabular}
  \caption{\textbf{Single vs. Multi-resolution in 3D Reconstruction.} The single-resolution baseline uses a fixed voxel size of 20 cm, while our multi-resolution setup dynamically adapts voxel sizes between 20 cm and 40 cm. Performance varies across sensor modalities due to data density. For \rgbd~input, the dense observations allow the multi-resolution approach to match the accuracy of the fixed-resolution baseline while reducing memory usage (\tabref{tab:memory_avg}). In contrast, for \lidar~input, sparser measurements limit the effectiveness of coarser voxels, leading to reduced reconstruction fidelity. This highlights the trade-off between adaptivity and resolution, especially in low-density regimes.}
  \label{tab:reconstruction_multi}
  
\end{table*}
}

\newcommand{\tablereplicaoxspiresreconstruction}{
\begin{table*}[htbp]
  \centering
  \footnotesize
  \setlength{\tabcolsep}{4pt}
  \renewcommand{\arraystretch}{1.4}
  \begin{tabular}{llcccccccc!{\vrule width 0.25pt}@{\hskip 4pt}ccc!{\vrule width 0.5pt}@{\hskip 6pt}ccc}
    \toprule
    & & \makecell{\tt{office0}} & \makecell{\tt{office1}} & \makecell{\tt{office2}} & \makecell{\tt{office3}} & \makecell{\tt{office4}} & \makecell{\tt{room0}} & \makecell{\tt{room1}} & \makecell{\tt{room2}} & \makecell{\tt{bod}} & \makecell{\tt{keb}} & \makecell{\tt{obs}} & \makecell{Avg\\{\tiny RGB-D}} & \makecell{Avg\\{\tiny LiDAR}} & \makecell{Avg} \\
    \midrule

    \rowcolor{colorSep}
    &&\multicolumn{8}{l}{\textit{\textbf{RGB-D}}}&\multicolumn{3}{l}{\textit{\textbf{LiDAR}}}& &\\

    \multirow{4}{*}{\rotatebox[origin=c]{90}{PIN-SLAM}}
    & Acc.[cm]$\downarrow$ & 0.946 & 0.821 & 0.885 & 1.041 & 0.871 & 0.861 & 0.851 & 0.872 & 8.672 & 7.681 & 9.365 & 0.893 & 8.573 & 4.733 \\
    & Comp.[cm]$\downarrow$ & 4.383 & 4.641 & \fs 3.332 & 2.686 & \fs 2.834 & 3.147 & 4.843 & \fs 2.560 & \fs 13.846 & \fs 10.579 & 12.577 & 3.553 & \fs 12.334 & \fs 7.944 \\
    & C-L1$\downarrow$ & 2.664 & 2.731 & 2.109 & 1.864 & 1.853 & 2.004 & 2.847 & 1.716 & \fs 11.259 & 9.130 & 10.971 & 2.223 & \nd 10.453 & 6.338 \\
    & F-score[\%]$\uparrow$ & 93.815 & 93.020 & \fs 94.604 & 95.370 & \fs 95.290 & 94.701 & 91.820 & \fs 96.121 & \fs 84.864 & \nd 89.193 & 85.378 & 94.343 & \fs 86.478 & \fs 90.411\\

    \midrule
    \multirow{4}{*}{\rotatebox[origin=c]{90}{VDBFusion}}
    & Acc.[cm]$\downarrow$ & \fs 0.545 & \fs 0.512 & \fs 0.554 & \fs 0.575 & \fs 0.568 & \fs 0.574 & \fs 0.535 & \fs 0.546 & \fs 5.447 & \nd 5.145 & \nd 6.951 & \fs 0.551 & \fs 5.848 & \fs 3.200 \\
    & Comp.[cm]$\downarrow$ & \nd 2.784 & \nd 3.717 & 3.498 & 2.693 & 2.892 & 2.539 & \nd 2.381 & \nd 2.647 & 21.987 & \nd 10.703 & \nd 12.125 & \nd 2.894 & 14.939 & \nd 8.917 \\
    & C-L1$\downarrow$ & \nd 1.664 & \nd 2.115 & \nd 2.026 & \nd 1.634 & \fs 1.730 & 1.557 & \nd 1.458 & \fs 1.597 & 13.717 & \fs 7.924 & \nd 9.538 & \nd 1.722 & \fs 10.393 & \fs 6.058\\
    & F-score[\%]$\uparrow$ & 95.828 & 93.804 & 94.115 & 94.916 & 94.853 & 95.685 & 95.773 & \nd 95.645 & 74.414 & \fs 90.519 & \nd 86.945 & 95.077 & 83.959 & \nd 89.518 \\

    \midrule
    \multirow{4}{*}{\rotatebox[origin=c]{90}{Voxblox}}
    & Acc.[cm]$\downarrow$ & \nd 0.577 & 0.682 & 0.699 & 0.896 & \nd 0.597 & 0.622 & 0.746 & \nd 0.574 & 8.484 & 8.772 & 9.001 & 0.674 & 8.753 & 4.714 \\
    & Comp.[cm]$\downarrow$ & 3.331 & 7.529 & 3.452 & \fs 2.434 & \nd 2.863 & \fs 2.184 & 3.116 & 2.826 & 21.861 & 23.340 & 18.994 & 3.467 & 21.398 & 12.433\\
    & C-L1$\downarrow$ & 1.954 & 4.106 & 2.075 & 1.665 & \fs 1.730 & \fs 1.403 & 1.931 & 1.700 & 15.172 & 16.056 & 13.998 & 2.070 & 15.076 &  8.573\\
    & F-score[\%]$\uparrow$ & 95.262 & 87.167 & 94.251 & \fs 95.706 & 94.817 & 96.223 & 94.647 & 94.997 & 73.172 & 70.284 & 76.241 & 94.134 & 73.233 & 83.684 \\

    \midrule
    \multirow{4}{*}{\rotatebox[origin=c]{90}{Supereight2$\dagger$}}
    & Acc.[cm]$\downarrow$ & 0.819 & 0.790 & 0.985 & 0.952 & 1.211 & 1.109 & 0.932 & 1.323 & \textcolor{red}{fail} & \textcolor{red}{fail} & \textcolor{red}{fail} & 1.015 & -- & -- \\
    & Comp.[cm]$\downarrow$ & 2.907 & 3.845 & 3.728 & 2.751 & 3.143 & 2.624 & 2.807 & 2.712 & \textcolor{red}{fail} & \textcolor{red}{fail} & \textcolor{red}{fail} & 3.065 & -- & -- \\
    & C-L1$\downarrow$ & 1.863 & 2.317 & 2.356 & 1.851 & 2.177 & 1.867 & 1.869 & 2.018 & \textcolor{red}{fail} & \textcolor{red}{fail} & \textcolor{red}{fail} & 2.040 & -- & -- \\
    & F-score[\%]$\uparrow$ & \fs 96.199 & \nd 93.998 & \nd 94.392 & \nd 95.594 & \nd 95.044 & \nd 96.321 & \nd 95.777 & 95.015 & \textcolor{red}{fail} & \textcolor{red}{fail} & \textcolor{red}{fail} & \fs 95.293 & -- & -- \\

    \midrule
    \multirow{4}{*}{\rotatebox[origin=c]{90}{Supereight2$\ddagger$}}
    & Acc.[cm]$\downarrow$ & 0.819 & 0.790 & 1.004 & 0.977 & 1.216 & 1.113 & 0.934 & 1.323 & \textcolor{red}{fail} & \textcolor{red}{fail} & \textcolor{red}{fail} & 1.022 & -- & -- \\
    & Comp.[cm]$\downarrow$ & 2.908 & 3.844 & 3.743 & 2.761 & 3.147 & 2.615 & 2.810 & 2.714 & \textcolor{red}{fail} & \textcolor{red}{fail} & \textcolor{red}{fail} & 3.068 & -- & -- \\
    & C-L1$\downarrow$ & 1.863 & 2.317 & 2.374 & 1.869 & 2.182 & 1.864 & 1.872 & 2.018 & \textcolor{red}{fail} & \textcolor{red}{fail} & \textcolor{red}{fail} & 2.045 & -- & -- \\
    & F-score[\%]$\uparrow$ & \nd 96.197 & \nd 93.998 & 94.384 & 95.586 & 95.039 & \fs 96.326 & 95.775 & 95.017 & \textcolor{red}{fail} & \textcolor{red}{fail} & \textcolor{red}{fail} & \nd 95.290 & -- & -- \\

    \midrule
    \multirow{4}{*}{\rotatebox[origin=c]{90}{\n3mapping}}
    & Acc.[cm]$\downarrow$ & \textcolor{red}{fail} & 0.739 & \textcolor{red}{fail} & \textcolor{red}{fail} & \textcolor{red}{fail} & \textcolor{red}{fail} & \textcolor{red}{fail} & \textcolor{red}{fail} & \nd 5.737 & 5.617 & \fs 6.649 & \textcolor{red}{0.739} & \nd 6.001 & -- \\
    & Comp.[cm]$\downarrow$ & \textcolor{red}{fail} & 3.950 & \textcolor{red}{fail} & \textcolor{red}{fail} & \textcolor{red}{fail} & \textcolor{red}{fail} & \textcolor{red}{fail} & \textcolor{red}{fail} & 25.728 & 13.379 & 12.389 & \textcolor{red}{3.950} & 17.165 & -- \\
    & C-L1$\downarrow$ & \textcolor{red}{fail} & 2.344 & \textcolor{red}{fail} & \textcolor{red}{fail} & \textcolor{red}{fail} & \textcolor{red}{fail} & \textcolor{red}{fail} & \textcolor{red}{fail} & 15.733 & 9.498 & \fs 9.519 & \textcolor{red}{2.344} & 11.583 & -- \\
    & F-score[\%]$\uparrow$ & \textcolor{red}{fail} & \fs 96.560 & \textcolor{red}{fail} & \textcolor{red}{fail} & \textcolor{red}{fail} & \textcolor{red}{fail} & \textcolor{red}{fail} & \textcolor{red}{fail} & 68.460 & 88.026 & \fs 88.458 & \textcolor{red}{96.560} & 81.648 & --\\

    \midrule
    \multirow{4}{*}{\rotatebox[origin=c]{90}{NKSR}}
    & Acc.[cm]$\downarrow$ & \textcolor{red}{fail} & \textcolor{red}{fail} & \textcolor{red}{fail} & \textcolor{red}{fail} & \textcolor{red}{fail} & \textcolor{red}{fail} & \textcolor{red}{fail} & \textcolor{red}{fail} & 8.782 & 7.654 & 8.798 & -- & 8.411 & -- \\
    & Comp.[cm]$\downarrow$ & \textcolor{red}{fail} & \textcolor{red}{fail} & \textcolor{red}{fail} & \textcolor{red}{fail} & \textcolor{red}{fail} & \textcolor{red}{fail} & \textcolor{red}{fail} & \textcolor{red}{fail} & \nd 15.804 & 11.318 & \fs 11.915 & -- & \nd 13.013 & -- \\
    & C-L1$\downarrow$ & \textcolor{red}{fail} & \textcolor{red}{fail} & \textcolor{red}{fail} & \textcolor{red}{fail} & \textcolor{red}{fail} & \textcolor{red}{fail} & \textcolor{red}{fail} & \textcolor{red}{fail} & \nd 12.293 & 9.486 & 10.357 & -- & 10.712 & -- \\
    & F-score[\%]$\uparrow$ & \textcolor{red}{fail} & \textcolor{red}{fail} & \textcolor{red}{fail} & \textcolor{red}{fail} & \textcolor{red}{fail} & \textcolor{red}{fail} & \textcolor{red}{fail} & \textcolor{red}{fail} & \nd 81.660 & 88.295 & 86.650 & -- & \nd 85.535 & -- \\

    \midrule
    \multirow{4}{*}{\rotatebox[origin=c]{90}{Ours}}
    & Acc.[cm]$\downarrow$ & 0.579 & \nd 0.542 & \nd 0.601 & \nd 0.624 & 0.624 & \nd 0.607 & \nd 0.565 & 0.594 & 6.293 & \fs 4.892 & 7.639 & \nd 0.592 & 6.275 & \nd 3.434 \\
    & Comp.[cm]$\downarrow$ & \fs 2.714 & \fs 3.685 & \nd 3.414 & \nd 2.582 & 2.880 & \nd 2.452 & \fs 2.311 & 2.704 & 21.269 & 12.698 & 12.743 & \fs 2.843 & 15.570 & 9.207\\
    & C-L1$\downarrow$ & \fs 1.646 & \fs 2.113 & \fs 2.007 & \fs 1.603 & \nd 1.752 & \nd 1.530 & \fs 1.438 & \nd 1.649 & 13.781 & \nd 8.795 & 10.191 & \fs 1.717 & 10.922 & \nd 6.320\\
    & F-score[\%]$\uparrow$ & 96.062 & 93.867 & 94.221 & 95.280 & 94.889 & 96.029 & \fs 96.012 & 95.601 & 75.065 & 88.462 & 85.865 & 95.245 & 83.131 & 89.188 \\

    \bottomrule
  \end{tabular}
  \caption{\textbf{3D Reconstruction Results}. The \rgbd~data are sequences of ScanNet~\cite{straub2019replica} and \lidar~data are sequences of Oxford Spires Dataset~\cite{tao2025oxford}. All the pipelines are run with ground-truth fixed poses. Voxel size is set to 1 cm for \rgbd~data and 20 cm for \lidar~data, and the F-score is computed with a 10 cm error threshold for the \rgbd~and 20 cm for the \lidar. Best results are highlighted as \colorbox{colorFst}{\textbf{first}} and \colorbox{colorSnd}{second}. Supereigth$\dagger$ represents single resolution grid, Supereight$\ddagger$ represents multi-resolution grid. Both Supereigth approaches run on Replica sequences but fail on Oxford Spires. \n3mapping fails on most Replica sequences but runs on Oxford. NKSR fails on all Replica sequences but runs on Oxford Spires.}
  \label{tab:replica_oxspires_reconstruction}
\end{table*}
}

\acmJournal{TOG}
\graphicspath{figures/}

\begin{document}
\title{Resolution Where It Counts: Hash-based GPU-Accelerated 3D Reconstruction via Variance-Adaptive Voxel Grids}

\author{Lorenzo De Rebotti}
\affiliation{%
 \institution{Sapienza, University of Rome}
 \city{Rome}
 \country{Italy}
}
\email{derebotti@diag.uniroma1.it}
\author{Emanuele Giacomini}
\affiliation{%
 \institution{Sapienza, University of Rome}
 \city{Rome}
 \country{Italy}
}
\email{giacomini@diag.uniroma1.it}
\author{Giorgio Grisetti}
\affiliation{%
 \institution{Sapienza, University of Rome}
 \city{Rome}
 \country{Italy}
}
\email{grisetti@diag.uniroma1.it}
\author{Luca Di Giammarino}
\affiliation{%
 \institution{Sapienza, University of Rome}
 \city{Rome}
 \country{Italy}
}
\email{digiammarino@diag.uniroma1.it}

\begin{abstract}
Efficient and scalable 3D surface reconstruction from range data remains a core challenge in computer graphics and vision, particularly in real-time and resource-constrained scenarios. Traditional volumetric methods based on fixed-resolution voxel grids or hierarchical structures like octrees often suffer from memory inefficiency, computational overhead, and a lack of GPU support.
We propose a novel variance-adaptive, multi-resolution voxel grid that dynamically adjusts voxel size based on the local variance of signed distance field (SDF) observations. Unlike prior multi-resolution approaches that rely on recursive octree structures, our method leverages a flat spatial hash table to store all voxel blocks, supporting constant-time access and full GPU parallelism. This design enables high memory efficiency and real-time scalability.
We further demonstrate how our representation supports GPU-accelerated rendering through a parallel quad-tree structure for Gaussian Splatting, enabling effective control over splat density. Our open-source CUDA/C++ implementation achieves up to 13× speedup and 4× lower memory usage compared to fixed-resolution baselines, while maintaining on par results in terms of reconstruction accuracy, offering a practical and extensible solution for high-performance 3D reconstruction.
\end{abstract}

%
%


%
%


\begin{teaserfigure}
\begin{flushleft}
\vspace{-0.3cm}
{\large \textcolor{magenta}{\texttt{\href{https://rvp-group.github.io/mrhash/}{https://rvp-group.github.io/mrhash/}}}}\\
\vspace{0.1cm}
\begin{center}
    \centering
    \small\sffamily
\begingroup%
  \makeatletter%
  \providecommand\color[2][]{%
    \errmessage{(Inkscape) Color is used for the text in Inkscape, but the package 'color.sty' is not loaded}%
    \renewcommand\color[2][]{}%
  }%
  \providecommand\transparent[1]{%
    \errmessage{(Inkscape) Transparency is used (non-zero) for the text in Inkscape, but the package 'transparent.sty' is not loaded}%
    \renewcommand\transparent[1]{}%
  }%
  \providecommand\rotatebox[2]{#2}%
  \newcommand*\fsize{\dimexpr\f@size pt\relax}%
  \newcommand*\lineheight[1]{\fontsize{\fsize}{#1\fsize}\selectfont}%
  \ifx\svgwidth\undefined%
    \setlength{\unitlength}{488.25621117bp}%
    \ifx\svgscale\undefined%
      \relax%
    \else%
      \setlength{\unitlength}{\unitlength * \real{\svgscale}}%
    \fi%
  \else%
    \setlength{\unitlength}{\svgwidth}%
  \fi%
  \global\let\svgwidth\undefined%
  \global\let\svgscale\undefined%
  \makeatother%
  \begin{picture}(1,0.28087732)%
    \lineheight{1}%
    \setlength\tabcolsep{0pt}%
    \put(0.63805937,0.26375388){\makebox(0,0)[t]{\lineheight{1.25}\smash{\begin{tabular}[t]{c}3DGS\end{tabular}}}}%
    \put(0,0){\includegraphics[width=\unitlength,page=1]{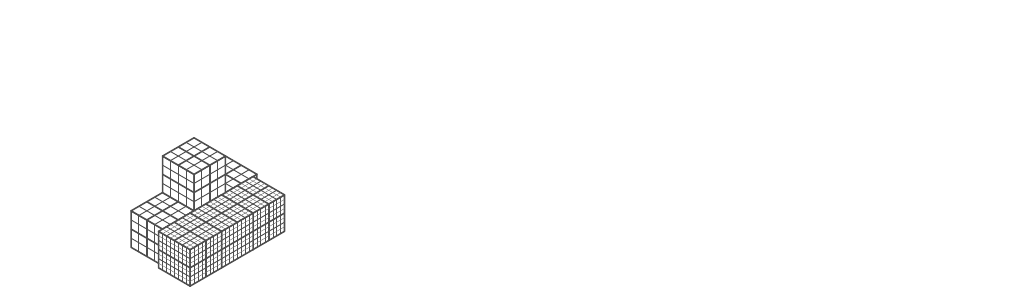}}%
    \put(0.18967574,0.26371645){\makebox(0,0)[t]{\lineheight{1.25}\smash{\begin{tabular}[t]{c}Sparse Multi \\Resolution Voxels\end{tabular}}}}%
    \put(0,0){\includegraphics[width=\unitlength,page=2]{figures/teaser_full.pdf}}%
    \put(0.41217099,0.26326185){\makebox(0,0)[t]{\lineheight{1.25}\smash{\begin{tabular}[t]{c}Surface Reconstruction\end{tabular}}}}%
    \put(0,0){\includegraphics[width=\unitlength,page=3]{figures/teaser_full.pdf}}%
    \put(0.78663944,0.26949315){\color[rgb]{0,0,0}\makebox(0,0)[lt]{\lineheight{1.25}\smash{\begin{tabular}[t]{l}F1\textsubscript{replica}\end{tabular}}}}%
    \put(0.92282603,0.26949315){\color[rgb]{0,0,0}\makebox(0,0)[lt]{\lineheight{1.25}\smash{\begin{tabular}[t]{l}F1\textsubscript{ncd}\end{tabular}}}}%
    \put(0.97775919,0.175359){\color[rgb]{0,0,0}\makebox(0,0)[lt]{\lineheight{1.25}\smash{\begin{tabular}[t]{l}F1\textsubscript{os}\end{tabular}}}}%
    \put(0.73729477,0.17537118){\color[rgb]{0,0,0}\makebox(0,0)[lt]{\lineheight{1.25}\smash{\begin{tabular}[t]{l}F1\textsubscript{SNet}\end{tabular}}}}%
    \put(0.79904234,0.07931){\color[rgb]{0,0,0}\makebox(0,0)[lt]{\lineheight{1.25}\smash{\begin{tabular}[t]{l}Memory\end{tabular}}}}%
    \put(0.9116961,0.07912252){\color[rgb]{0,0,0}\makebox(0,0)[lt]{\lineheight{1.25}\smash{\begin{tabular}[t]{l}FPS\end{tabular}}}}%
    \put(0,0){\includegraphics[width=\unitlength,page=4]{figures/teaser_full.pdf}}%
    \put(0.04002826,0.26417848){\makebox(0,0)[t]{\lineheight{1.25}\smash{\begin{tabular}[t]{c}RGB-D\\LiDAR\end{tabular}}}}%
    \put(0,0){\includegraphics[width=\unitlength,page=5]{figures/teaser_full.pdf}}%
    \put(0.8312355,0.02514079){\makebox(0,0)[lt]{\lineheight{1.25}\smash{\begin{tabular}[t]{l}voxblox\end{tabular}}}}%
    \put(0,0){\includegraphics[width=\unitlength,page=6]{figures/teaser_full.pdf}}%
    \put(0.8312355,0.04374787){\makebox(0,0)[lt]{\lineheight{1.25}\smash{\begin{tabular}[t]{l}vdbfusion\end{tabular}}}}%
    \put(0,0){\includegraphics[width=\unitlength,page=7]{figures/teaser_full.pdf}}%
    \put(0.8312355,0.06235495){\makebox(0,0)[lt]{\lineheight{1.25}\smash{\begin{tabular}[t]{l}ours\end{tabular}}}}%
    \put(0,0){\includegraphics[width=\unitlength,page=8]{figures/teaser_full.pdf}}%
    \put(0.8312355,0.00723016){\makebox(0,0)[lt]{\lineheight{1.25}\smash{\begin{tabular}[t]{l}pin-slam\end{tabular}}}}%
  \end{picture}%
\endgroup%

    \caption{\textbf{Overview}. Our pipeline for real-time 3D reconstruction and rendering. \textit{Left}: Our variance-adaptive voxel grid dynamically adjusts resolution based on TSDF variance, enabling compact and accurate reconstructions from both \rgbd~and unstructured 3D point clouds. The mesh output supports high-quality rendering via Gaussian Splatting, with adaptive splat control handled fully on the GPU. \textit{Right}: Overview comparison reporting F-score across different datasets, memory efficiency, and FPS.}
    \Description{System Overview}
\end{center}
\end{flushleft}
\end{teaserfigure}

\maketitle
\section{Introduction}

Accurate and scalable 3D reconstruction \cite{digiammarino2023photometric,giacomini2025splatloam} from depth or point cloud data is a core challenge in computer graphics, robotics, and vision. Applications such as robotic navigation, AR/VR, and large-scale mapping demand representations that are both geometrically precise and computationally efficient. Volumetric methods, particularly the \ac{tsdf}, are widely used for this purpose.
Introduced by Curless~and~Levoy~\cite{curless1996volumetric}, \ac{tsdf}-based fusion integrates depth over time into a voxel grid, enabling robust surface reconstruction.
Traditional approaches \cite{newcombe2011kinectfusion,niessner2013real,oleynikova2017voxblox} employ uniform grids, which scale poorly due to high memory and compute demands. 

To improve efficiency, adaptive structures like octrees and OpenVDB~\cite{museth2013vdb} have been proposed. Octrees hierarchically allocate memory where needed but suffer from recursive traversal and poor GPU performance. OpenVDB provides sparse multi-resolution storage through a tree structure, and variants such as VDBFusion~\cite{vizzo2022vdbfusion} have extended it to \ac{tsdf} fusion. Despite their versatility, these structures introduce non-trivial access overhead and are not optimized for real-time, GPU-based reconstruction. Furthermore, many existing multi-resolution approaches depend on input-specific cues (e.g., semantics or depth confidence), limiting their generality and adaptability across sensor types. 

In this work, we propose \our, a novel variance-adaptive, multi-resolution voxel grid implemented entirely on the GPU using a flat spatial hash table. Our system adjusts resolution based on local \ac{tsdf} variance, independent of sensor modality or semantic priors, enabling fine detail where needed and coarse representation elsewhere. Unlike octrees, our approach avoids hierarchical traversal, achieving, on average, constant-time access and real-time performance.

Our main contributions are:
\begin{itemize}
\item A variance-adaptive voxel grid that adjusts resolution based on local \ac{tsdf} variance, preserving fine detail and reducing memory in uniform areas. We extend Marching Cubes to support seamless meshing across resolution boundaries.
\item A flat hash table structure that manages mixed-resolution voxel blocks without hierarchical overhead, enabling efficient GPU integration and access.
\item A fully parallel GPU-based quad-tree for controlling the spatial distribution of Gaussians for novel view synthesis, allowing adaptive splat density without CPU-GPU streaming.
\item An open-source CUDA/C++ implementation supporting real-time reconstruction and rendering, designed for extensibility and community use.
\end{itemize}
\section{Related Work}
\label{sec:related}
3D mapping systems typically represent geometry as either explicit sparse maps \cite{brizi2024vbr,cwian2025madba} or implicit volumetric fields\cite{niessner2013real}. Explicit pipelines maintain sparse local submaps composed of samples or parametric primitives (e.g., point clouds, surfels, or 2D/3D Gaussian splats) that are well-suited for scan-to-scan \cite{digiammarino2022mdslam} or scan-to-submap registration and odometry \cite{ferrari2024madicp,zhang2014loam,schops2019bad}. In contrast, implicit volumetric approaches store a discretized signed distance field, enabling robust multi-view fusion, denoising, and direct meshing at the cost of higher memory and compute.

Volumetric 3D reconstruction using the \ac{tsdf} has been a foundational approach for over three decades. The seminal work by Curless and Levoy~\cite{curless1996volumetric} introduced volumetric fusion of depth maps, enabling robust surface reconstruction through SDF accumulation. Although initially restricted to offline processing, this approach laid the groundwork for real-time systems.
KinectFusion~\cite{newcombe2011kinectfusion} demonstrated real-time TSDF integration on a dense, uniform voxel grid, enabling accurate reconstructions in small, bounded scenes. However, the fixed resolution and exhaustive memory allocation made the approach unsuitable for large-scale environments. Voxel Hashing~\cite{niessner2013real} addressed this limitation by dynamically allocating voxel blocks using a spatial hash table, reducing memory usage and enabling real-time mapping for mobile platforms.

Building on this foundation, works like Voxblox~\cite{oleynikova2017voxblox} and Supereight~\cite{vespa2019adaptive} further improved scalability. Voxblox uses block-wise hashing and ray bundling to support onboard updates on resource-constrained UAVs, while Supereight introduces an adaptive allocation strategy to improve runtime and memory efficiency. These systems prioritize real-time performance, often at the cost of geometric precision.

OpenVDB~\cite{museth2013vdb} and the newer deep learning framework~\cite{williams2024fvdb} offer a more general-purpose sparse volumetric data structure widely used in graphics and VFX. OpenVDB supports multi-resolution storage through a hierarchical tree structure and has been adapted to 3D reconstruction in works like VDBFusion~\cite{vizzo2022vdbfusion}, which applies it to unstructured 3D point cloud fusion. However, OpenVDB’s multi-level hierarchy introduces non-trivial lookup costs and is not well-suited for real-time integration. Additionally, each leaf node operates at a fixed resolution, limiting local adaptivity within a region.

In parallel, learning-based representations have emerged that model 3D geometry via continuous neural fields. Methods such as InstantNGP~\cite{muller2022instant}, NICER-SLAM~\cite{zhu2024nicerslam}, and PIN-SLAM~\cite{pan2024pinslam} leverage hierarchical feature grids and MLP decoding to reconstruct high-quality geometry while incorporating SLAM constraints. Although these approaches achieve accurate results, they are typically limited to small-scale scenes and require substantial computational and memory resources. 
Fully neural pipelines such as NKSR~\cite{huang2023neural} and \n3mapping~\cite{song2024n3mapping} achieve high fidelity even under sparse observations; however, both depend on computationally intensive per-scene neural optimization and demand high GPU resources, often leading to failures when scaling to large-scale scenarios.

Recent works have explored adaptive resolution guided by semantics or image gradients~\cite{zheng2024mapadapt}, enabling selective detail refinement. However, such approaches depend on structured \rgbd~input and do not generalize to unstructured 3D point clouds. Similarly, Funk et al.~\cite{funk2021multiresolution} use octree structures for adaptive occupancy mapping, but rely on log-odds surface modeling and suffer from hierarchical access overhead.

More recently, 3D Gaussian Splatting~\cite{kerbl20233d} has been combined with~\ac{tsdf} mapping, as in GSFusion~\cite{wei2024gsfusion}. They extend Supereight2~\cite{funk2021multiresolution} to support online rendering. The \ac{tsdf} layer provides stable geometry for constraining Gaussians initialization and optimization.

Unlike prior work, we introduce a variance-driven, multi-resolution voxel grid that is sensor-agnostic and GPU-native. Instead of relying on semantic cues or hierarchical traversal, our system adaptively allocates voxel resolution based on local \ac{tsdf} variance, using a flat hash table. This enables constant-time access, resolution mixing, and real-time performance across both \rgbd~and \lidar~inputs, without sacrificing geometric fidelity or memory efficiency.
\section{Technical Section}
\begin{figure*}[ht]
    \centering
    \small\sffamily
    \resizebox{\linewidth}{!}{%
    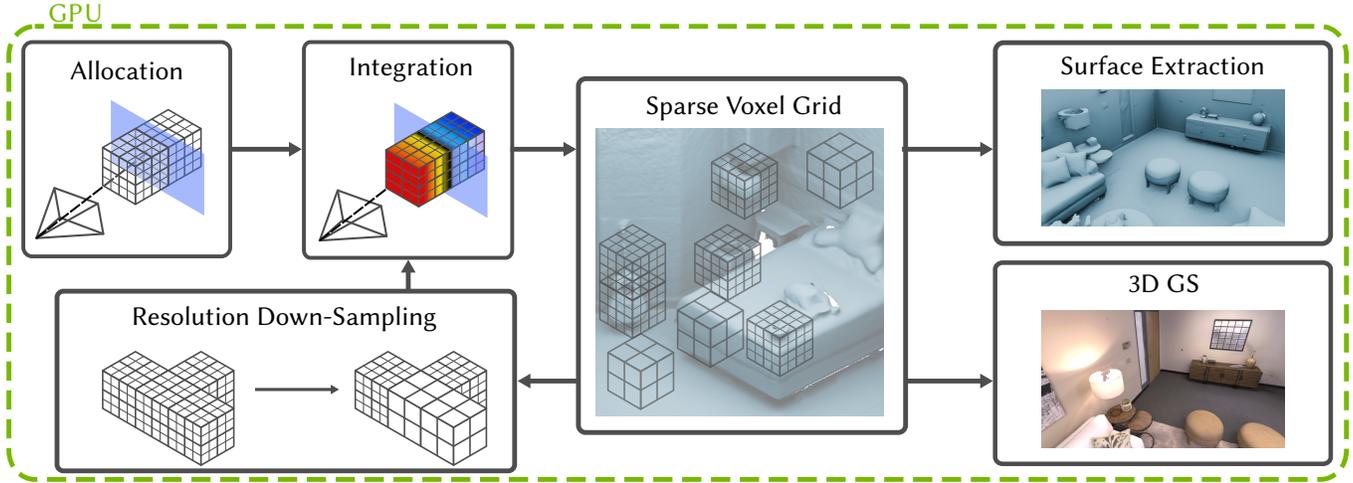
    }
    \caption{\textbf{System Overview:} \our~takes the input data and, if necessary, allocates the voxel blocks associated with such 3D points and integrates them into a multi-resolution sparse voxel grid. The blocks whose variance is below a predefined threshold are downsampled and re-integrated into the voxel grid at a coarser resolution. Moreover, our pipeline can either extract the isosurface from the volumetric representation or render the surrounding environment.}
    \Description{}
    \label{fig:pipeline}
\end{figure*}
The \ac{tsdf} map representation used consists of a collection of spatially hashed voxels $\bm{V}_i$ with a variable voxel size $ \nu \in \mathbb{R}^+$, which is determined based on the density of neighboring voxels. Voxels of the same size are stored in $\bm{B}_j$ blocks (cuboids), each of which contains a number of voxels that varies with the voxel resolution. 
Each block $\bm{B}_j$ is identified by a global position $\mathbf{b}_j = \left (x, y, z \right)^T \in \mathbb{Z}^3$. 
A voxel $\bm{V}_i$ mainly stores a truncated signed distance $D_i \in \mathbb{R}$, a weight $W_i \in \mathbb{N}$ that reflects the confidence in $D_i$, the color, and the variance of the mean \ac{tsdf} $\sigma_i^2$.

At each frame $k$, the system integrates the current sensor's measurement (either a raw point cloud or a dense image).

\subsection{Integration}
\label{sec:integration}
We integrate 3D sensor data into a \ac{tsdf} volume using a standard volumetric fusion strategy~\cite{curless1996volumetric}. The \ac{tsdf} provides an implicit surface representation, where each voxel $\bm{V}_i$ encodes a scalar value $D_i$ indicating the signed distance from the voxel center to the closest observed surface along the sensor’s line of sight. Positive values represent free space (outside the surface), while negative values correspond to space behind the surface (inside objects). To restrict updates to relevant regions, the distance is truncated beyond a fixed threshold $\tau > 0$. Depending on the type of measurement and the availability of dense color data, we distinguish between different integration logic.

\subsubsection{3D Point Cloud.} For 3D point cloud measurements (\ie,~\lidar~ sensors), we perform ray-based geometric integration using a DDA traversal~\cite{amanatides1987fast} from the sensor origin to each point $\mathbf{p}$. Voxels intersected by the ray are updated based on their signed distance to $\mathbf{p}$. For a voxel center at position $\mathbf{x}$ and a surface point $\mathbf{p}$ observed by the sensor, the truncated signed distance is defined as:

\begin{equation}
d_k(\mathbf{x}) = \text{clip}\left( \left( \mathbf{p} - \mathbf{x} \right) \cdot \hat{\mathbf{n}}, -\tau, \tau \right),
\end{equation}   
where $\hat{\mathbf{n}}=\frac{\bp-\bo}{\lVert\bp-\bo\rVert}$ is the unit ray direction from the sensor origin $\bo$ to the point $\bp$.

\subsubsection{Cameras.} Differently, for cameras with providing depth information(\ie,~\rgbd), we use a projective mapping approach. Each voxel center $\mathbf{x}$ is projected onto the image plane to obtain pixel coordinates $(u,v)$. The corresponding depth value $d$ at that pixel defines the measured distance along the viewing ray. The signed distance is then computed as the difference between the measured depth and the voxel depth $\|\mathbf{x}\|$ along the same ray:

\begin{equation}
d_k(\mathbf{x}) = \text{clip}\left( d - \| \mathbf{x} \|, -\tau, \tau \right).
\end{equation}

We note that the above integration strategy follows standard volumetric fusion techniques~\cite{curless1996volumetric, niessner2013real}, and is not a core contribution of our work. However, we include it here for completeness and clarity of exposition.

Independent of the integration type, each voxel maintains a weighted average of these observations; the distance $D_i$ and the weight $W_i$ are updated as follows:

\begin{equation}
\label{eq:update}
D_i \leftarrow \frac{W_i D_i + w_k d_k(\mathbf{x})}{W_i + w_k}, \quad
W_i \leftarrow W_i + w_k.
\end{equation}

Here, $w_k$ is a confidence weight. While it is often set based on sensor noise or distance \cite{niessner2013real, curless1996volumetric}, we fix $w_k = 1$ to simplify the computation of voxel-wise variance. Different from other works, with the running average $D_{i}$, we maintain the sample variance $\sigma_{i}^2$ of the \ac{tsdf} values using an online update. This variance reflects the local geometric complexity: low in flat regions, high near surface boundaries, or noise. We use $\sigma_{i}^2$ to guide voxel resolution, preserving fine detail where needed and merging voxels in smoother areas, enabling adaptive resolution and improved memory efficiency.

\begin{figure}[htbp]
    \centering
    \small\sffamily
    \resizebox{\linewidth}{!}{%
    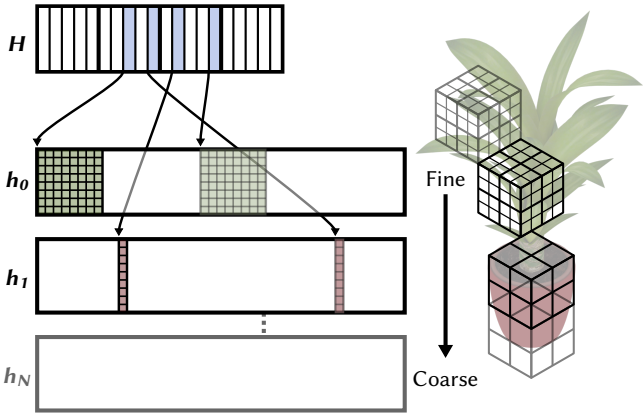
    }
    \caption{\textbf{Illustration of grid representation to hash table mapping}. The hash function maps the world points from integer world coordinates to the respective buckets of the hash-table $\bm{H}$. Every entry of the hash-table contains a pointer to a block in the specific heap of that resolution.} 
    \label{fig:svo}
    \Description{}
\end{figure}

\subsection{Grid Representation}
\label{sec:tech-grid}
Given the inherent sparsity of most 3D environments, a dense volumetric grid would be highly impractical for large-scale scenarios. Instead, we adopt a sparse volumetric representation based on a hash table, as introduced in \cite{niessner2013real}, which enables memory-efficient storage and access to active voxel blocks. While this approach introduces additional challenges related to data locality and memory coherence, it allows for efficient large-scale scene reconstruction without requiring predefined spatial bounds. One of our main contributions is a multi-resolution grid tailored for a single hash table. Unlike fixed-voxel-size methods \cite{oleynikova2017voxblox, niessner2013real}, we allocate voxel blocks of a constant metric size, while varying their internal resolution by adjusting the number of voxels contained within each block. Each entry in the hash table $\bm{H}$ contains:
\begin{itemize}
\item $\mathbf{b}_j \in \mathbb{Z}^3$: integer coordinates of the voxel block origin (lower-left corner), used as the spatial key;
\item $o_j \in \mathbb{N}$: auxiliary offset for resolving hash collisions;
\item $t_j \in \mathbb{Z}$: pointer to the voxel block data stored in GPU memory;
\item $n_j \in \mathbb{N}$: index of the resolution level corresponding to heap $\bm{h}_n$.
\end{itemize}
Each hash entry uniquely identifies a voxel block $\bm{B}_j$ through its spatial key $\mathbf{b}_j$, which is mapped to an index by the hash function $H$, defined as:

\begin{equation}
\label{eq:hashing}
H(x,y,z)=\left(x \cdot p_{1} \oplus y \cdot p_{2} \oplus z \cdot p_{3}\right)\operatorname{mod} n_{hash}
\end{equation}
where \(p_{1} = 73856093, p_{2}=19349669, p_{3}=83492791\) are large prime numbers~\cite{teschner2003optimized} and $n_{\textrm{hash}}$ is the number of entries in the hash table.

As in \cite{niessner2013real}, to handle collisions, each entry in the hash table maintains a fixed-size bucket for up to $N$ voxel blocks. In the rare event of overflow, additional blocks are appended to a per-entry linked list. In addition, as is commonly done \cite{niessner2013real,oleynikova2017voxblox} to allocate all intersected voxel blocks efficiently, we leverage DDA~\cite{amanatides1987fast}. 

The key idea of our method is having a unified hash table supporting a multi-resolution voxel representation within a single address space.
Each entry of the hash table stores a pointer to the voxel block allocated in heap $\bm{h}_n$ associated with that resolution level.
This structure allows voxel blocks of different resolutions to be indexed and accessed through the same hash table (see \figref{fig:svo}).

While managing multiple resolutions within a flat hash table increases implementation complexity (e.g., in hashing, memory layout, and consistency), it enables a more flexible and efficient system compared to traditional hierarchical representations \cite{vespa2018efficient}.
Most multi-resolution volumetric approaches use octree hierarchies \cite{funk2021multiresolution}, which naturally support dyadic spatial subdivision and adaptive refinement near surfaces. 
However, octrees incur higher computational cost, typically $O(\mathrm{log}N)$, for insertion, deletion, and neighborhood queries due to recursive traversal and pointer-based memory layouts. 
Even methods leveraging hybrid tree-based data structures, such as those proposed in~\cite{museth2013vdb, museth2021nanovdb}, provide highly efficient sparse volumetric representations; however, their hierarchical organization introduces considerable overhead for frequent updates and random access.
In contrast, our method achieves constant-time access $O(1)$ on average for these operations by encoding multi-resolution voxel blocks directly into a flat hash table. 

Resolution is handled explicitly through key encoding, enabling the simultaneous representation of voxels at different scales within a unified, non-hierarchical structure. This avoids the overhead of recursive descent and facilitates more efficient parallel execution on modern hardware, particularly in GPU-based implementations where memory indirection and dynamic branching are costly. We show the benefits of our implementation in \secref{sec:experiments}.

\subsubsection{Voxel Merging Criterion}
\label{sec:tech-variance}
\begin{figure}[htbp]
    \label{fig:merging}
    \centering
    \small\sffamily
    \resizebox{\linewidth}{!}{%
\begingroup%
  \makeatletter%
  \providecommand\color[2][]{%
    \errmessage{(Inkscape) Color is used for the text in Inkscape, but the package 'color.sty' is not loaded}%
    \renewcommand\color[2][]{}%
  }%
  \providecommand\transparent[1]{%
    \errmessage{(Inkscape) Transparency is used (non-zero) for the text in Inkscape, but the package 'transparent.sty' is not loaded}%
    \renewcommand\transparent[1]{}%
  }%
  \providecommand\rotatebox[2]{#2}%
  \newcommand*\fsize{\dimexpr\f@size pt\relax}%
  \newcommand*\lineheight[1]{\fontsize{\fsize}{#1\fsize}\selectfont}%
  \ifx\svgwidth\undefined%
    \setlength{\unitlength}{210.78120434bp}%
    \ifx\svgscale\undefined%
      \relax%
    \else%
      \setlength{\unitlength}{\unitlength * \real{\svgscale}}%
    \fi%
  \else%
    \setlength{\unitlength}{\svgwidth}%
  \fi%
  \global\let\svgwidth\undefined%
  \global\let\svgscale\undefined%
  \makeatother%
  \begin{picture}(1,0.28680667)%
    \lineheight{1}%
    \setlength\tabcolsep{0pt}%
    \put(0,0){\includegraphics[width=\unitlength,page=1]{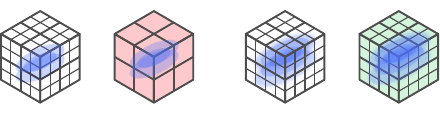}}%
    \put(0.21297952,0.00871479){\color[rgb]{0,0,0}\makebox(0,0)[t]{\lineheight{1.25}\smash{\begin{tabular}[t]{c}(a)\end{tabular}}}}%
    \put(0.76999971,0.00871771){\color[rgb]{0,0,0}\makebox(0,0)[t]{\lineheight{1.25}\smash{\begin{tabular}[t]{c}(b)\end{tabular}}}}%
    \put(0,0){\includegraphics[width=\unitlength,page=2]{figures/merging_strategy.pdf}}%
  \end{picture}%
\endgroup%

    }
    \caption{\textbf{Adopted merging strategy.} a) Voxel block with low \ac{tsdf} variance is reallocated at a coarser resolution (red). b) Voxel block with high \ac{tsdf} variance remains unchanged (green).}
    \Description{}
    \label{fig:svo_merging}
\end{figure}

We initialize the voxel blocks to the finest resolution, since ``ideally'' this provides the best geometrical accuracy. Then, to preserve surface complexity and handle high-frequency signals, we maintain higher voxel resolution in areas with high \ac{tsdf} variance over time. Instead, for regions exhibiting low variance, we merge voxels of the same block, as in \figref{fig:svo_merging}. 

Driven exclusively by \ac{tsdf} statistics, the approach is sensor-agnostic, requiring no sensor-specific assumptions or noise parameters.
We compute variance using a numerically stable, single-pass method based on Welford's algorithm \cite{welford1962note}. 

Given a sequence of $k$ observations $d_1,d_2,\dots,d_k$ for a voxel $\bm{V}_i$, we maintain the running mean, as defined in \eqref{eq:update}, and the accumulated sum of squared deviations $S_{2,i,k}$, updated as:

\begin{equation}
S_{2,i,k} = S_{2,i,k-1} + (d_k - D_{i,k-1})(d_k - D_{i,k}),
\end{equation}
from which the variance $\sigma_i^2$ is computed as:
\begin{equation}
\sigma_i^2 = \frac{S_{2,i,k}}{k}.
\end{equation}

This formulation enables incremental computation of voxel-wise variance without storing the full observation history.
At each integration step, variance is updated for all voxels and aggregated at the block level to obtain the mean variance of each block.
Because this operation is computationally demanding, we implement it in a fully parallel manner on the GPU, using intra-block reductions to efficiently accumulate the $S_{2,i,k}$ terms across voxels.

\subsection{Streaming}
\label{sec:tech-streaming}
GPU memory capacity and bandwidth are fundamental bottlenecks in large-scale 3D reconstruction, where volumetric data scales with the environment's size. Retaining all voxel data in GPU memory becomes infeasible for expansive scenes, and frequent CPU-GPU transfers introduce non-negligible latency due to limited bandwidth and synchronization costs. To address this, it is essential to adopt a streaming strategy that efficiently manages data movement. 

Unlike previous systems such as \cite{niessner2013real}, which continuously stream voxel blocks based on geometric heuristics (e.g., frustum culling or proximity thresholds), our method defers streaming until GPU memory usage approaches a user-defined capacity threshold (e.g., 85\% of the total GPU memory allocated for voxel data). Once this limit is reached, blocks are selected for removal from the GPU memory based on spatial relevance: for \rgbd~input, we consider blocks outside the current camera frustum; for \lidar~input, we consider blocks lying beyond a predefined radius from the current sensor pose. This deferred, capacity-driven policy minimizes unnecessary data transfers and synchronization overhead, maintaining GPU resources for actively reconstructed regions.

\subsection{Multi-resolution Marching Cubes}
\begin{figure}[htbp]
    \centering
    \small\sffamily
    \resizebox{0.9\linewidth}{!}{%
\begingroup%
  \makeatletter%
  \providecommand\color[2][]{%
    \errmessage{(Inkscape) Color is used for the text in Inkscape, but the package 'color.sty' is not loaded}%
    \renewcommand\color[2][]{}%
  }%
  \providecommand\transparent[1]{%
    \errmessage{(Inkscape) Transparency is used (non-zero) for the text in Inkscape, but the package 'transparent.sty' is not loaded}%
    \renewcommand\transparent[1]{}%
  }%
  \providecommand\rotatebox[2]{#2}%
  \newcommand*\fsize{\dimexpr\f@size pt\relax}%
  \newcommand*\lineheight[1]{\fontsize{\fsize}{#1\fsize}\selectfont}%
  \ifx\svgwidth\undefined%
    \setlength{\unitlength}{193.07719073bp}%
    \ifx\svgscale\undefined%
      \relax%
    \else%
      \setlength{\unitlength}{\unitlength * \real{\svgscale}}%
    \fi%
  \else%
    \setlength{\unitlength}{\svgwidth}%
  \fi%
  \global\let\svgwidth\undefined%
  \global\let\svgscale\undefined%
  \makeatother%
  \begin{picture}(1,0.27601036)%
    \lineheight{1}%
    \setlength\tabcolsep{0pt}%
    \put(0,0){\includegraphics[width=\unitlength,page=1]{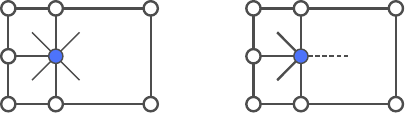}}%
    \put(0.21725971,0.1807386){\color[rgb]{0,0,0}\makebox(0,0)[t]{\lineheight{1.25}\smash{\begin{tabular}[t]{c}\textbf{?}\end{tabular}}}}%
    \put(0.21725971,0.0650003){\color[rgb]{0,0,0}\makebox(0,0)[t]{\lineheight{1.25}\smash{\begin{tabular}[t]{c}\textbf{?}\end{tabular}}}}%
    \put(0,0){\includegraphics[width=\unitlength,page=2]{figures/transvoxels.pdf}}%
  \end{picture}%
\endgroup%

    }
    \caption{\textbf{Ambiguity of \ac{sdf} interpolation.} The blue vertex attempts to interpolate the \ac{sdf} value from its four neighboring voxels during sampling, but encounters undefined entries due to missing neighbors. To address this, we apply a weighted interpolation scheme that prioritizes finer-resolution values where available. For clarity, the visualization is shown in 2D assuming bilinear interpolation.}
    \label{fig:multi-res-issue}
    \Description{}
\end{figure}

\begin{figure}[htbp]
    \centering
    \small\sffamily
    \resizebox{0.9\linewidth}{!}{%
\begingroup%
  \makeatletter%
  \providecommand\color[2][]{%
    \errmessage{(Inkscape) Color is used for the text in Inkscape, but the package 'color.sty' is not loaded}%
    \renewcommand\color[2][]{}%
  }%
  \providecommand\transparent[1]{%
    \errmessage{(Inkscape) Transparency is used (non-zero) for the text in Inkscape, but the package 'transparent.sty' is not loaded}%
    \renewcommand\transparent[1]{}%
  }%
  \providecommand\rotatebox[2]{#2}%
  \newcommand*\fsize{\dimexpr\f@size pt\relax}%
  \newcommand*\lineheight[1]{\fontsize{\fsize}{#1\fsize}\selectfont}%
  \ifx\svgwidth\undefined%
    \setlength{\unitlength}{174.04724842bp}%
    \ifx\svgscale\undefined%
      \relax%
    \else%
      \setlength{\unitlength}{\unitlength * \real{\svgscale}}%
    \fi%
  \else%
    \setlength{\unitlength}{\svgwidth}%
  \fi%
  \global\let\svgwidth\undefined%
  \global\let\svgscale\undefined%
  \makeatother%
  \begin{picture}(1,0.40176875)%
    \lineheight{1}%
    \setlength\tabcolsep{0pt}%
    \put(0,0){\includegraphics[width=\unitlength,page=1]{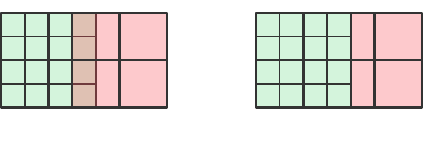}}%
    \put(0.24246115,0.01054884){\color[rgb]{0,0,0}\makebox(0,0)[t]{\lineheight{1.25}\smash{\begin{tabular}[t]{c}(a)\end{tabular}}}}%
    \put(0.78501627,0.01054983){\color[rgb]{0,0,0}\makebox(0,0)[t]{\lineheight{1.25}\smash{\begin{tabular}[t]{c}(b)\end{tabular}}}}%
    \put(0,0){\includegraphics[width=\unitlength,page=2]{figures/interpolation_problem.pdf}}%
  \end{picture}%
\endgroup%

    }
    \caption{\textbf{Transitional voxels.} (a) Due to the measurement-driven allocation, overlapping regions (dotted area) can occur between voxel blocks of different resolutions, which poses challenges during the vertex evaluation step in Marching Cubes. (b) To address this, we reduce the size of coarser voxels and adjust the placement of Marching Cubes vertices accordingly to maintain consistency across resolutions.}
    \label{fig:transvoxel}
    \Description{}
\end{figure}

To extract the isosurface from the \ac{tsdf}, we implement the Marching Cubes algorithm~\cite{lorensen1987marching} directly on the multi-resolution voxel grid, leveraging GPU parallelism. Each voxel is processed independently, using local signed distance values that are tri-linearly interpolated at its corners. Although voxel size does not affect the core algorithm, resolution boundaries pose a challenge: interpolating corner values involves querying neighboring voxels, whose \ac{tsdf} values are defined relative to different centers (see \figref{fig:multi-res-issue}).
Zheng et al. \cite{zheng2024mapadapt} address this by uniformly refining neighboring voxels, which increases memory usage in regions that would otherwise remain coarse. Instead, we assume limited variation at the scales considered and interpolate between coarse and fine voxels using a weighted scheme that favors the finer resolution where available.
Unlike voxel-based terrain rendering in computer graphics, such as Transvoxel \cite{lengyel2010transition}, a volumetric integration system like ours does not allow arbitrary allocation of voxel blocks, as their creation is driven by incoming measurements and does not account for neighboring voxel resolutions. Hence, outer coarse voxels may spatially overlap with adjacent finer-resolution voxels. To resolve this, inspired by \cite{lengyel2010transition},  we truncate coarse voxels along the dimensions corresponding to faces shared with finer neighbors, ensuring a clean partitioning of space and preventing multiple faces from being generated within the same spatial region (see \figref{fig:transvoxel}).
To further mitigate artifacts from these approximations, we apply a vertex collapsing step (based on voxel size) during mesh generation, simplifying redundant geometry and improving consistency across resolution transitions.

\subsection{Rendering}
\label{sec:tech-rendering}
We employ 3D Gaussian Splatting~\cite{kerbl20233d} for the \ac{nvs} task.
In this representation, the scene is modeled as a collection of $M$ 3D Gaussian primitives, each parametrized by a mean position $\bmu_{m,\scaleto{W}{3.6pt}}$ and a covariance matrix $\bSigma_{m,\scaleto{W}{3.6pt}}$, both defined in world coordinates, along with a view-dependent color $\mathbf{C}_m$, and an opacity $\alpha_m$.

Each 3D Gaussian in world coordinates is projected onto the image plane via a projective transformation $\pi$:

\begin{equation}
    \bmu_{m,\scaleto{I}{3.6pt}} = \pi\left(\mathbf{T}_{\scaleto{CW}{3.6pt}}\bmu_{m,\scaleto{W}{3.6pt}}\right), \quad \bSigma_{m,\scaleto{I}{3.6pt}} = \mathbf{JR}\bSigma_{m,\scaleto{W}{3.6pt}}\mathbf{R}^T\mathbf{J}^T
\end{equation}
where $\mathbf{T}_{\scaleto{CW}{3.6pt}} \in \mathbb{SE}(3)$ is the world-to-camera transform matrix, $\mathbf{J}$ is the Jacobian of the projective transformation $\pi$, and $\mathbf{R} \in \mathbb{SO}(3)$ is the rotation component of $\mathbf{T}_{\scaleto{CW}{3.6pt}}$. 
Since the covariance is defined in image space, the resulting $\bSigma_{m,\scaleto{I}{3.6pt}}$ is obtained by taking the upper-left $2\times2$ submatrix of the projected $3\times3$ covariance.
This formulation is fully differentiable; the means, covariances, colors, and opacities of the Gaussian can be jointly optimized from a set of posed input images using Stochastic Gradient Descent. 
The optimization process minimizes the photometric difference between the reference views and the corresponding rendered images. These are generated by a dedicated tile-based rasterizer that performs efficient $\alpha$-blending composing of the $M$ primitives in screen space. For each pixel $(u, v)$, the final color is computed as:

\begin{equation}
\label{eq:alpha-blend}
\mathbf{C}_{p}(u, v) = \sum_{m=1}^{M} \mathbf{C}_m \alpha_m \mathcal{G}_{m,\scaleto{I}{3.6pt}}(u,v)\prod^{m-1}_{l=1}\left(1-\alpha_{l}\mathcal{G}_{l,\scaleto{I}{4pt}}(u,v)\right) 
\end{equation}
where the primitives are indexed by depth, from front to back, and $\mathcal{G}_{m,\scaleto{I}{3.6pt}}(u,v)$ is the evaluation of the projected $m$-th Gaussian at the corresponding pixel, computed as
\begin{equation}
    \mathcal{G}_{m,\scaleto{I}{3.6pt}}(u,v) = \exp\left(-\frac{1}{2}\left((u,v) - \bmu_{m,\scaleto{I}{3.6pt}}\right)^{T} \bSigma_{m,\scaleto{I}{3.6pt}}^{-1}\left((u,v) - \bmu_{m,\scaleto{I}{3.6pt}}\right)\right)
\end{equation}

During optimization, the number of Gaussians is managed by an adaptive control mechanism that, based on gradient information, either prunes or densifies the primitives within specific regions.
Adaptive control of Gaussian density is key to an efficient representation, and the original implementation has several critical flaws. 
First, some areas may be oversaturated by primitives, increasing the rendering times, memory consumption, and training time.
Moreover, densification stems from existing Gaussians, so initial seeding is crucial to ensure fast convergence.

Following the new trend in 3DGS-based SLAM \cite{keetha2024splatam,matsuki2024gaussian,peng2024rtgslam}, we reformulate 3DGS in an incremental manner, processing temporally-ordered posed images sequentially.

The Gaussians are seeded using depth information, effectively solving the second problem, while our underlying voxel grid enables a more structured, efficient approach to controlling Gaussians and addresses oversaturation. To regulate Gaussian density, we construct an image-space quadtree guided by visibility and density cues derived from the voxel grid. Each node represents an image region whose contrast, \ie, the luminance-weighted RGB mean-squared error, quantifies local intensity variation and determines whether further subdivision is required.

To estimate the contrast $c$ of a quad-tree $\bm{Q}$, we consider the RGB values $\mathbf{q} \in \mathbb{R}^{3}$ for every pixel of the quad-tree and compute the following:
\begin{equation}
\label{eq:contrast}
c(\bm{Q}) = \mathbf{l} \ \frac{\sum_{\mathbf{q}\in \bm{Q}}  (\mathbf{q} - \operatorname{mean}_{\mathrm{rgb}}(\bm{Q}))^{2}}{\lvert \bm{Q} \lvert} 
\end{equation}
where $\mathbf{l} = [0.2989, 0.5870, 0.1140]$ are the Rec. 601 luma weights, $\operatorname{mean}_{\mathrm{rgb}}(\bm{Q}) \in \mathbb{R}^{3}$ are the mean values for the three channels of the quadtree's pixels, and $\lvert \bm{Q} \lvert$ is the number of pixels of the quadtree.
Within this context, $(\mathbf{q} - \operatorname{mean}_{\mathrm{rgb}}(\bm{Q}))^{2}$ denotes the element-wise square of the difference vector.

Similar to \cite{wei2024gsfusion}, the quadtrees are constructed by this contrast-based approach, dividing the image so that each leaf node contains a subregion with contrast below a threshold or a minimum pixel size. Unlike them, we construct quadtrees on the GPU and implement the operations in \eqref{eq:contrast} using the sum-reduction technique.

To maintain high throughput, both contrast estimation and quadtree construction are executed in parallel using a breadth-first strategy, avoiding recursion and leveraging shared memory for reductions (see \algref{alg:quadtree_subdivision}).
Therefore, all supporting data already resides in GPU memory, eliminating costly transfers and accelerating the auxiliary computations that guide adaptive splat generation.

\begin{algorithm}[htbp]
\caption{GPU-Based Quadtree Subdivision (Parallel Version)}
\label{alg:quadtree_subdivision}
\SetAlgoLined
\SetKwProg{Fn}{Procedure}{}{}
\SetKwInOut{Input}{Data}
\SetKwInOut{Output}{Result}
\SetKwFunction{ComputeContrast}{ComputeContrastWSharedMem}
\SetKwFunction{Subdivide}{Subdivide}

\Input{Input image, contrast threshold, minimum pixel size}
\Output{Final set of quadtree leaves}

Initialize \textit{RootNode} covering the entire image\;
$\mathit{NodeQueue}\gets\{\textit{RootNode}\}$\;

\While{$\mathit{NodeQueue}\neq\varnothing$}{
  \ForEach(\tcp*[f]{executed in parallel}){$\mathit{Node}\in\mathit{NodeQueue}$}{
      $\mathit{Node.Contrast}\gets$\ComputeContrast(\textit{Node}, Image)\;
  }

  \BlankLine
  $\mathit{NewQueue}\gets\varnothing$\;

  \ForEach(\tcp*[f]{executed in parallel}){$\mathit{Node}\in\mathit{NodeQueue}$}{
      \If{$\mathit{Node.Contrast}>\mathit{Threshold}$ \textbf{and} $\mathit{Node.Size}>\mathit{MinPixelSize}$}{
          $\mathit{Children}\gets$\Subdivide(\textit{Node})\;
          add $\mathit{Children}$ to $\mathit{NewQueue}$\;
      }\Else{
          add \textit{Node} to \textit{FinalLeaves}\;
      }
  }
  $\mathit{NodeQueue}\gets\mathit{NewQueue}$\;
} 
\Return \textit{FinalLeaves}\;

\BlankLine
\BlankLine
\Fn{\textsc{Subdivide}(Node)}{%
  Compute midpoints in $x$ and $y$\;%
  Create 4 children: TopLeft, BottomLeft, TopRight, BottomRight\;%
  \Return Children\;%
}%
\end{algorithm}
\section{Experiments}
\label{sec:experiments}
In this section, we report the results of our method on several publicly available datasets. We evaluate \our~ for both mapping accuracy and rendering performance.
To the best of our knowledge, ours is the first approach to combine a variance-driven, multi-resolution voxel grid with a single flat hash table, enabling fully GPU-based \ac{tsdf} integration and rendering.

\subsection{Experimental Settings}
\label{sec:exp-settings}
\subsubsection{Datasets.}
\label{sec:exp-datasets}
To evaluate the performance of our method, we used several publicly available benchmarks that feature heterogeneous sequences and varying resolutions across different sensor types.
\rgbd~Replica Dataset~\cite{straub2019replica} is a synthetic benchmark consisting of 18 photorealistic, semantically annotated indoor environments at room scale. Each scene provides dense meshes textured with high-resolution HDR imagery, per-primitive semantic class and instance annotations, as well as explicit modeling of reflective surfaces such as glass and mirrors. \rgbd~images are rendered from ground-truth poses using the official renderer.
Scannet~\cite{dai2017scannet} was constructed from 2.5M \rgbd~frames across 1513 indoor scenes, recorded using iPad-mounted depth sensors. The raw video streams were processed with BundleFusion~\cite{dai2017bundlefusion} to yield globally consistent meshes, which were then densely annotated via large-scale crowdsourcing into semantic and instance-level segmentations. Ground truth includes camera intrinsics, 6-DoF poses, and reconstructed surfaces.
\ac{nc}~\cite{zhang2021multicamera} was collected using a handheld Ouster OS0-128 \lidar~across both structured environments and vegetated areas. Ground truth was obtained with a Leica BLK360 scanner, providing centimeter-level accuracy for both poses and map points.
Oxford Spires Dataset~\cite{tao2025oxford} was recorded using a handheld Hesai QT64 \lidar~sensor, which provides a 360° horizontal field of view, 104° vertical field of view, 64 vertical channels, and a maximum range of 60 meters. Each sequence is accompanied by a survey-grade 3D laser scan, serving as a prior map for ground-truth trajectory estimation and mapping evaluation.

\subsubsection{Evaluation Metrics.}
\label{sec:exp-metrics}
To evaluate mapping quality, we adopt standard surface reconstruction metrics as defined in \cite{mescheder2019occupancy}. Accuracy (Acc) measures the average distance from points on the reconstructed mesh to their nearest neighbors in the reference point cloud, while Completeness (Comp) captures the inverse, how well the reconstruction covers the reference surface. The Chamfer-$L_1$ distance (C-$L_1$) is the mean of these two and provides a balanced measure of geometric fidelity. Additionally, we compute the F-score with a 10 cm error threshold for the \rgbd~and 20 cm for the \lidar. F-score evaluates geometric consistency by combining precision and recall into a single value.
The rendering quality is measured comparing \ac{psnr}, \ac{ssim}~\cite{wang2004image}, and \ac{lpips}~\cite{zhang2018unreasonable}.

\subsubsection{Baselines.}
\label{sec:exp-baselines}
We compared our method against popular mapping SOTA pipelines.
We include VDBFusion~\cite{vizzo2022vdbfusion}, VoxBlox~\cite{oleynikova2017voxblox}, and Supereight2 \cite{funk2021multiresolution} as purely geometric baselines. VDBFusion employs OpenVDB~\cite{museth2013vdb}  as a volumetric data structure to handle 3D points, VoxBlox combines adaptive weights and grouped ray-casting for an efficient \ac{tsdf} integration, while Supereight2 supports a multi-resolution grid through an octree implementation.
Additionally, we include two neural-implicit mapping pipelines. \n3mapping~\cite{song2024n3mapping} is a neural-implicit non-projective \ac{sdf} mapping pipeline. It leverages normal guidance to produce more accurate \ac{sdf}s, leading to SOTA results for offline \lidar~mapping. PIN-SLAM~\cite{pan2024pinslam}, leverages neural points as primitives and interleaves an incremental learning of the model's \ac{sdf} and, using marching cubes, it can produce a mesh from the underlying implicit \ac{sdf}.
In addition, we compare our method with other approaches that incorporated Gaussian Splatting into their map representation, like \cite{wei2024gsfusion}. Different from us, they employed an octree-based map representation, which aids the splats initialization.
For both 3D reconstruction and rendering experiments, we ran all the pipelines with ground-truth poses.

\subsubsection{Implementation.}
\label{sec:exp-implementation}
The experiments were run on a PC with an Intel Core i9-13900K @ 3.20Ghz, 128GB of RAM, and an NVIDIA RTX 4090 GPU with 24 GB of VRAM.
For all experiments where we employed our single-resolution configuration, we used a block size of 512 voxels. Also, for the finer resolution of the multi-resolution grid, we use a block size of 512; consequently, for the coarser resolution, there are 64 blocks. For the hash table, we employed a bucket size of 10 hash entries, an eventual linked list size of 7 elements, and, as presented in \secref{sec:tech-grid}, the values used to compute the hash of the block coordinates are: \(p_{1} = 73856093, p_{2}=19349669, p_{3}=83492791\).
In the rendering experiments, all the approaches used the same parameters to subdivide the image into quadtrees: a threshold contrast of 0.1 and a minimum pixel size of 1.
To optimize the Gaussian parameters, we used the same solver configuration for all methods, except for Ours (+iterations), where we applied our single-resolution grid and increased the number of iterations in the Gaussian Splatting solver.

Our results show that our method (supporting both \rgbd~and point cloud inputs) offers improved performance and efficiency while maintaining a SOTA accuracy in 3D reconstruction.

\tablefinalreconstruction
\tablereplicaoxspiresreconstruction

\subsection{3D Reconstruction}
\begin{figure*}[htbp]
    \centering
    \small\sffamily
    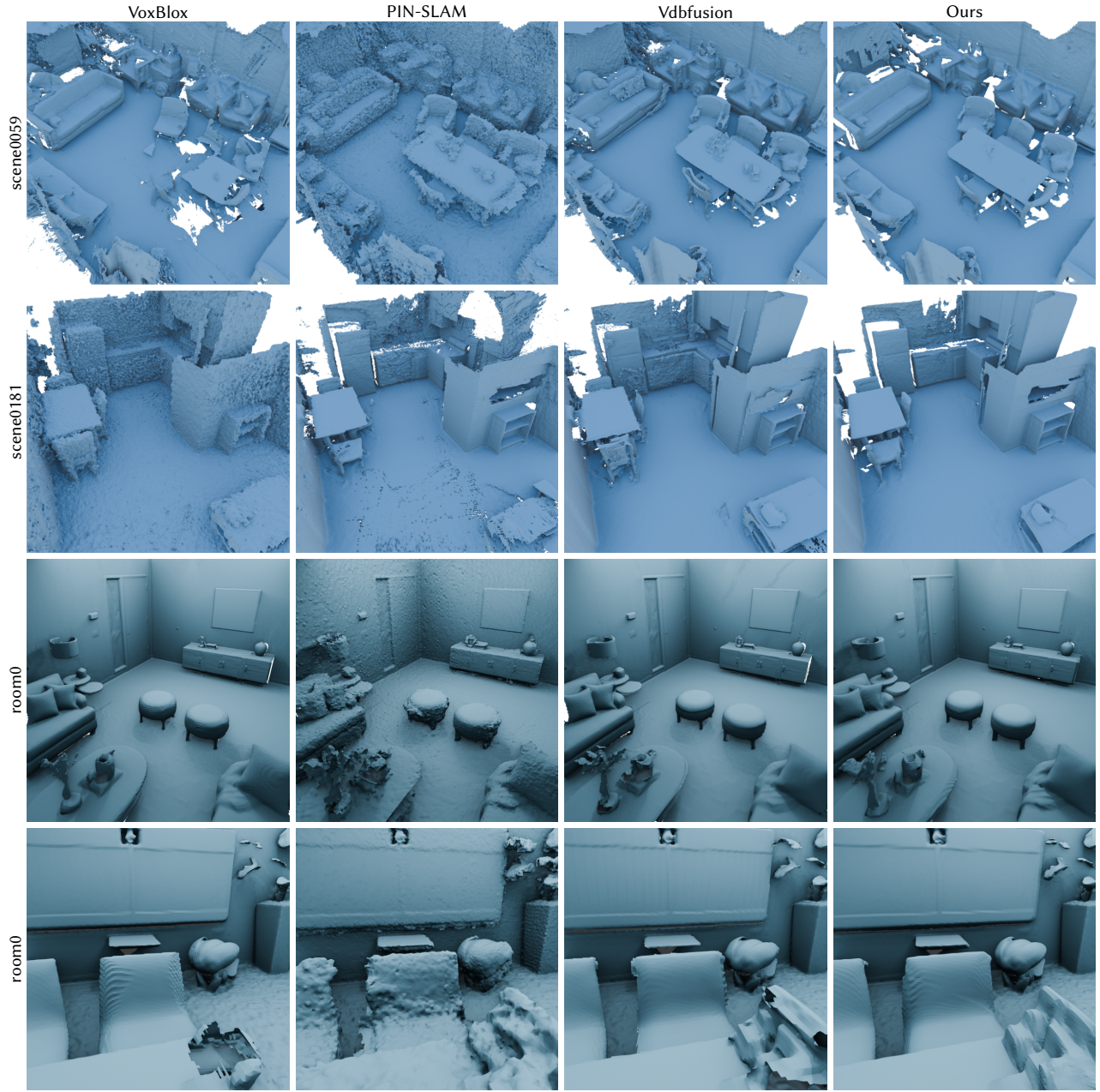
    \caption{\textbf{Qualitative results on \rgbd~Datasets.} The images show the mapping results on scene0059 and scene0181 of ScanNet dataset~\cite{dai2017scannet} and sequence room0 of Replica dataset \cite{straub2019replica}. We compare our method against other SOTA.}
    \label{fig:qualitative_rgbd}
    \Description{}
\end{figure*}

\begin{figure*}[htbp]
    \centering
    \small\sffamily
    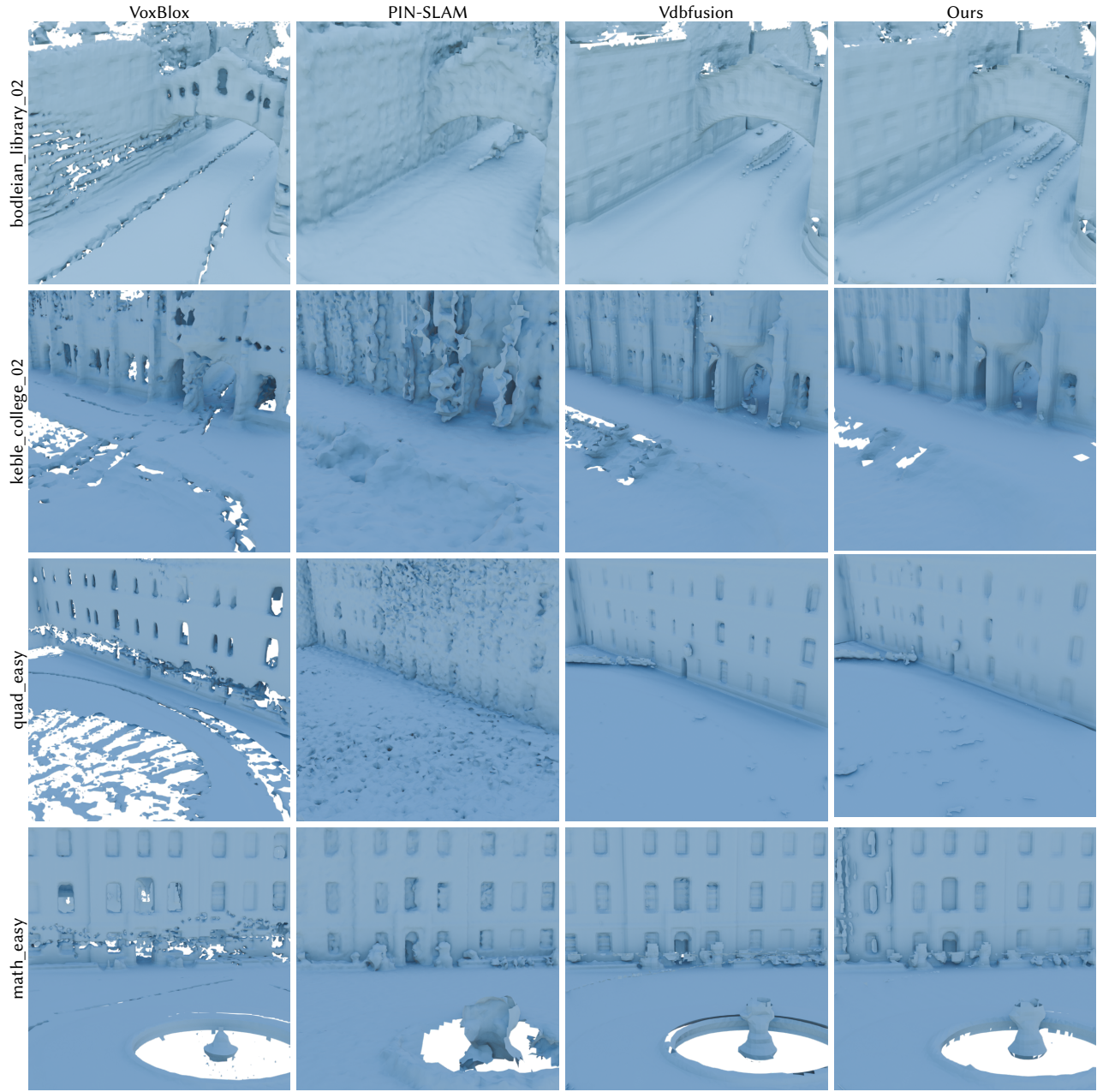
    \caption{\textbf{Qualitative results on \lidar~ Datasets.} The images show the mapping results on sequences bodleian-library-02 and keble-college-02 Oxford-Spires~\cite{tao2025oxford} and sequences quad-easy and math-easy of \ac{nc}~\cite{zhang2021multicamera}. We compare our method against other SOTA.}
    \label{fig:qualitative_lidar}
    \Description{}
\end{figure*}

Our method achieves reconstruction quality comparable to or better than prior approaches across multiple datasets. The results demonstrate that our variance-adaptive multi-resolution grid does not compromise accuracy, even while significantly reducing memory usage (\tabref{tab:memory_avg}). Notably, \n3mapping failed to complete any sequence from the \ac{nc} dataset, likely due to its high computational and memory requirements, which limit its scalability. For Voxblox, we used the ``FastTSDF'' mode for \rgbd~input, as the standard configuration was unstable or slow in our tests. For \lidar~data, we switched to the ``MergedTSDF'' configuration to ensure compatibility with sparser point clouds. These settings were chosen to reflect best-case performance for each method in its intended domain. To ensure a fair comparison, voxel sizes were adapted to sensor resolution: 0.01 m for dense \rgbd~input and 0.20 m for sparse \lidar~input. This allows each method to operate under conditions that match the characteristics of its input, and helps isolate the effect of voxel structure and memory management. \tabref{tab:final_reconstruction} and \tabref{tab:replica_oxspires_reconstruction} present the quantitative results on Scannet with \ac{nc}, as well as Replica combined with Oxford-Spires, respectively. Complementary qualitative examples are provided in \figref{fig:qualitative_rgbd} and \figref{fig:qualitative_lidar}. These visuals illustrate how our multi-resolution approach maintains surface detail in geometrically complex regions and avoids over-allocation in homogeneous areas, supporting both high accuracy and efficiency across sensor types.

\tablememory

\subsection{Memory}
We report memory consumption indirectly by measuring the number of mesh vertices and faces produced during reconstruction, which closely correlates with storage requirements and runtime memory usage. \tabref{tab:runtimes} summarizes average values across both \rgbd~and \lidar~datasets. Our method, particularly in its multi-resolution configuration, produces significantly more compact reconstructions compared to existing approaches. In the \rgbd~setup, we observe memory savings ranging from 2.0× to 7.5×. For \lidar~data, reductions range from 2.3× to 2.9× compared to other baselines, except for VDBFusion, whose mesh size is 0.7x of ours. 
On average, our multi-resolution compression method improves from 1.95x to 4.07x.
This demonstrates the effectiveness of our variance-driven voxel allocation, which allows the system to retain high resolution in geometrically complex areas while using coarser voxels in flat or redundant regions.  These gains are made possible by our design, which avoids over-allocating memory in uniform areas through adaptive resolution. 

\subsection{Runtimes}
\tabref{tab:runtimes} reports both the processing time per frame (in milliseconds) and the corresponding \ac{fps}. Our approach outperforms all baselines in nearly every scenario. The single-resolution variant already yields significant speedups, up to 13× faster than prior state-of-the-art methods such as VDBFusion and Voxblox. The multi-resolution configuration achieves additional performance gains, especially in large-scale or sparse scenes, while preserving reconstruction quality. Notably, our method maintains real-time or near-real-time performance across all datasets, including challenging \lidar-based sequences from \ac{nc}. Competing methods such as \n3mapping failed to complete these sequences, highlighting the scalability and robustness of our approach. These improvements stem from three key design choices: (1) a flat hash table that enables constant-time access without hierarchical traversal overhead, (2) a fully GPU-based pipeline that avoids costly CPU-GPU data transfers, and (3) an adaptive voxel resolution strategy that reduces unnecessary computation and memory usage in low-detail regions. Together, these contribute to the significant runtime advantage of our system.

\tableruntimes

\begin{figure}[htbp]
    \centering
    \small\sffamily
    \includegraphics[width=0.96\linewidth]{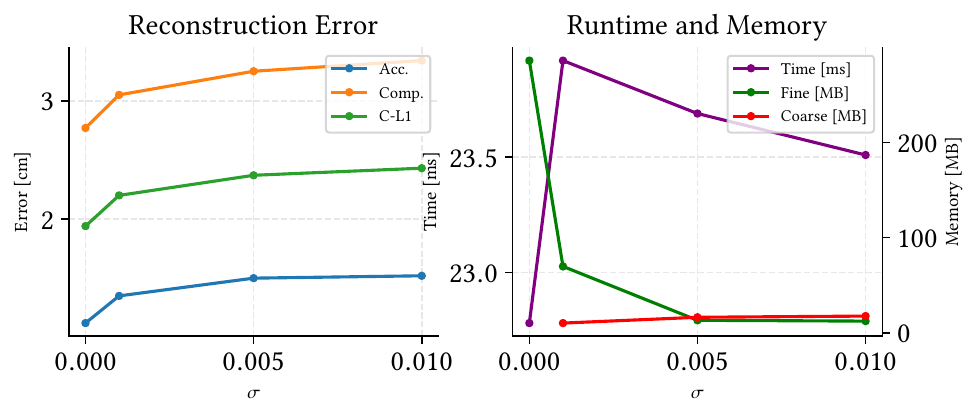}
    \caption{\textbf{Comparison of reconstruction error, runtimes, and memory usage for different variance thresholds $\sigma$.} The left plot reports accuracy, completeness, and Chamfer distance, while the right plot shows runtime and memory consumption of the fine- and coarse-level voxel structures. The results confirm that low thresholds yield the highest accuracy at the cost of increased fine-resolution memory, while moderate thresholds preserve comparable quality with reduced memory usage.}
    \label{fig:ablation_threshold}
    \Description{}
\end{figure}

\tablegswithtiming
\tablegswrecon

\tablemultimethod
\begin{figure*}[htbp]
    \centering
    \small\sffamily
    \resizebox{0.76\linewidth}{!}{%
\begingroup%
  \makeatletter%
  \providecommand\color[2][]{%
    \errmessage{(Inkscape) Color is used for the text in Inkscape, but the package 'color.sty' is not loaded}%
    \renewcommand\color[2][]{}%
  }%
  \providecommand\transparent[1]{%
    \errmessage{(Inkscape) Transparency is used (non-zero) for the text in Inkscape, but the package 'transparent.sty' is not loaded}%
    \renewcommand\transparent[1]{}%
  }%
  \providecommand\rotatebox[2]{#2}%
  \newcommand*\fsize{\dimexpr\f@size pt\relax}%
  \newcommand*\lineheight[1]{\fontsize{\fsize}{#1\fsize}\selectfont}%
  \ifx\svgwidth\undefined%
    \setlength{\unitlength}{439.89603839bp}%
    \ifx\svgscale\undefined%
      \relax%
    \else%
      \setlength{\unitlength}{\unitlength * \real{\svgscale}}%
    \fi%
  \else%
    \setlength{\unitlength}{\svgwidth}%
  \fi%
  \global\let\svgwidth\undefined%
  \global\let\svgscale\undefined%
  \makeatother%
  \begin{picture}(1,0.26744573)%
    \lineheight{1}%
    \setlength\tabcolsep{0pt}%
    \put(0,0){\includegraphics[width=\unitlength,page=1]{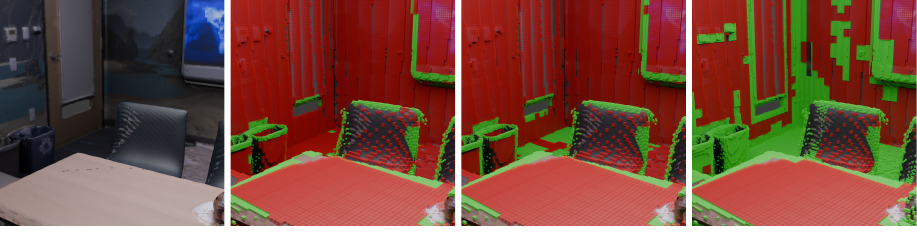}}%
    \put(0.12242258,0.0048651){\color[rgb]{0,0,0}\makebox(0,0)[t]{\lineheight{1.25}\smash{\begin{tabular}[t]{c}reference\end{tabular}}}}%
    \put(0.3741275,0.0048651){\color[rgb]{0,0,0}\makebox(0,0)[t]{\lineheight{1.25}\smash{\begin{tabular}[t]{c}$\sigma = 0.01$\end{tabular}}}}%
    \put(0.62583243,0.0048651){\color[rgb]{0,0,0}\makebox(0,0)[t]{\lineheight{1.25}\smash{\begin{tabular}[t]{c}$\sigma = 0.005$\end{tabular}}}}%
    \put(0.8775374,0.0048651){\color[rgb]{0,0,0}\makebox(0,0)[t]{\lineheight{1.25}\smash{\begin{tabular}[t]{c}$\sigma = 0.001$\end{tabular}}}}%
  \end{picture}%
\endgroup%
}
    \caption{\textbf{Comparison of different variance thresholds $\sigma$.} In red, the coarser voxels, in green, the finer ones. From left to right: the reference ground-truth mesh and the underlying grid at the respective thresholds. For simplicity, we show just two levels at a time. The sequence is part of the Replica dataset \cite{straub2019replica}.}
    \label{fig:res_comparison}
    \Description{}
\end{figure*}

\subsection{Rendering}
\begin{figure*}[htbp]
    \centering
    \small\sffamily
\begingroup%
\makeatletter%
\providecommand\color[2][]{%
	\errmessage{(Inkscape) Color is used for the text in Inkscape, but the package 'color.sty' is not loaded}%
	\renewcommand\color[2][]{}%
}%
\providecommand\transparent[1]{%
	\errmessage{(Inkscape) Transparency is used (non-zero) for the text in Inkscape, but the package 'transparent.sty' is not loaded}%
	\renewcommand\transparent[1]{}%
}%
\providecommand\rotatebox[2]{#2}%
\newcommand*\fsize{\dimexpr\f@size pt\relax}%
\newcommand*\lineheight[1]{\fontsize{\fsize}{#1\fsize}\selectfont}%
\ifx\svgwidth\undefined%
	\setlength{\unitlength}{507.10203552bp}%
	\ifx\svgscale\undefined%
		\relax%
	\else%
		\setlength{\unitlength}{\unitlength * \real{\svgscale}}%
	\fi%
\else%
	\setlength{\unitlength}{\svgwidth}%
\fi%
\global\let\svgwidth\undefined%
\global\let\svgscale\undefined%
\makeatother%
\begin{picture}(1,0.58177837)%
	\lineheight{1}%
	\setlength\tabcolsep{0pt}%
	\put(0,0){\includegraphics[width=\unitlength,page=1]{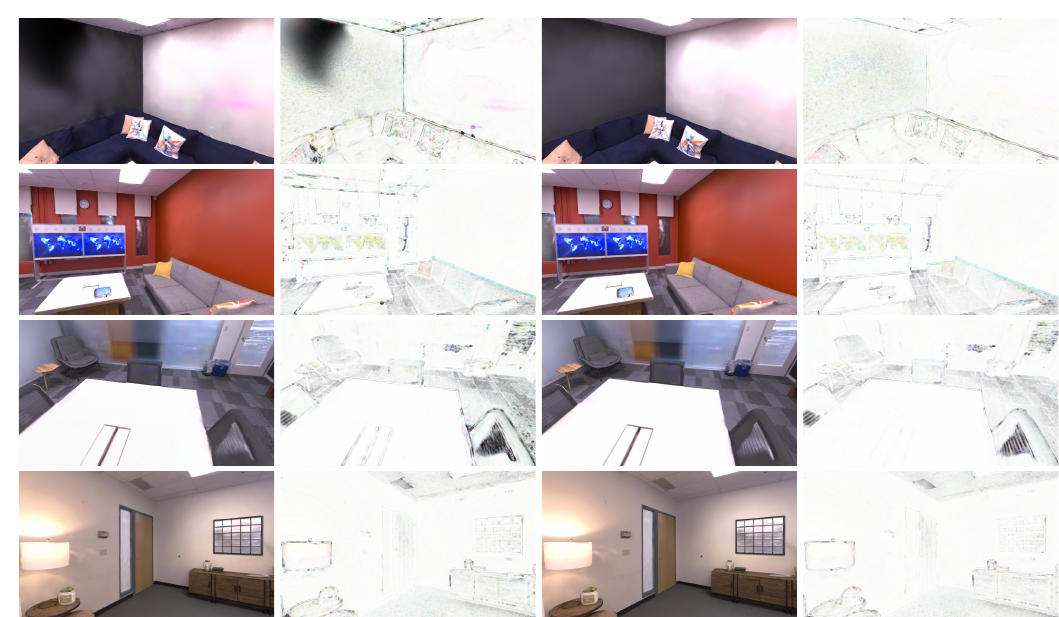}}%
	\put(0.26197959,0.56714622){\color[rgb]{0,0,0}\makebox(0,0)[t]{\lineheight{1.25}\smash{\begin{tabular}[t]{c}GSFusion\end{tabular}}}}%
	\put(0.75596524,0.56714622){\color[rgb]{0,0,0}\makebox(0,0)[t]{\lineheight{1.25}\smash{\begin{tabular}[t]{c}Ours\end{tabular}}}}%
	\put(0.01498706,0.4958832){\color[rgb]{0,0,0}\rotatebox{90}{\makebox(0,0)[t]{\lineheight{1.25}\smash{\begin{tabular}[t]{c}office2\end{tabular}}}}}%
	\put(0.01498706,0.35335716){\color[rgb]{0,0,0}\rotatebox{90}{\makebox(0,0)[t]{\lineheight{1.25}\smash{\begin{tabular}[t]{c}office3\end{tabular}}}}}%
	\put(0.01498706,0.21083111){\color[rgb]{0,0,0}\rotatebox{90}{\makebox(0,0)[t]{\lineheight{1.25}\smash{\begin{tabular}[t]{c}office4\end{tabular}}}}}%
	\put(0.01498706,0.06830507){\color[rgb]{0,0,0}\rotatebox{90}{\makebox(0,0)[t]{\lineheight{1.25}\smash{\begin{tabular}[t]{c}room0\end{tabular}}}}}%
\end{picture}%
\endgroup%

    \caption{\textbf{Qualitative \ac{nvs} results on \rgbd~Datasets.} Column 1 and 3 shows rendered RGB frames while column 2 and 4 shows the corresponding \ac{ssim} error.}
    \label{fig:qualitative_rendering}
    \Description{}
\end{figure*}
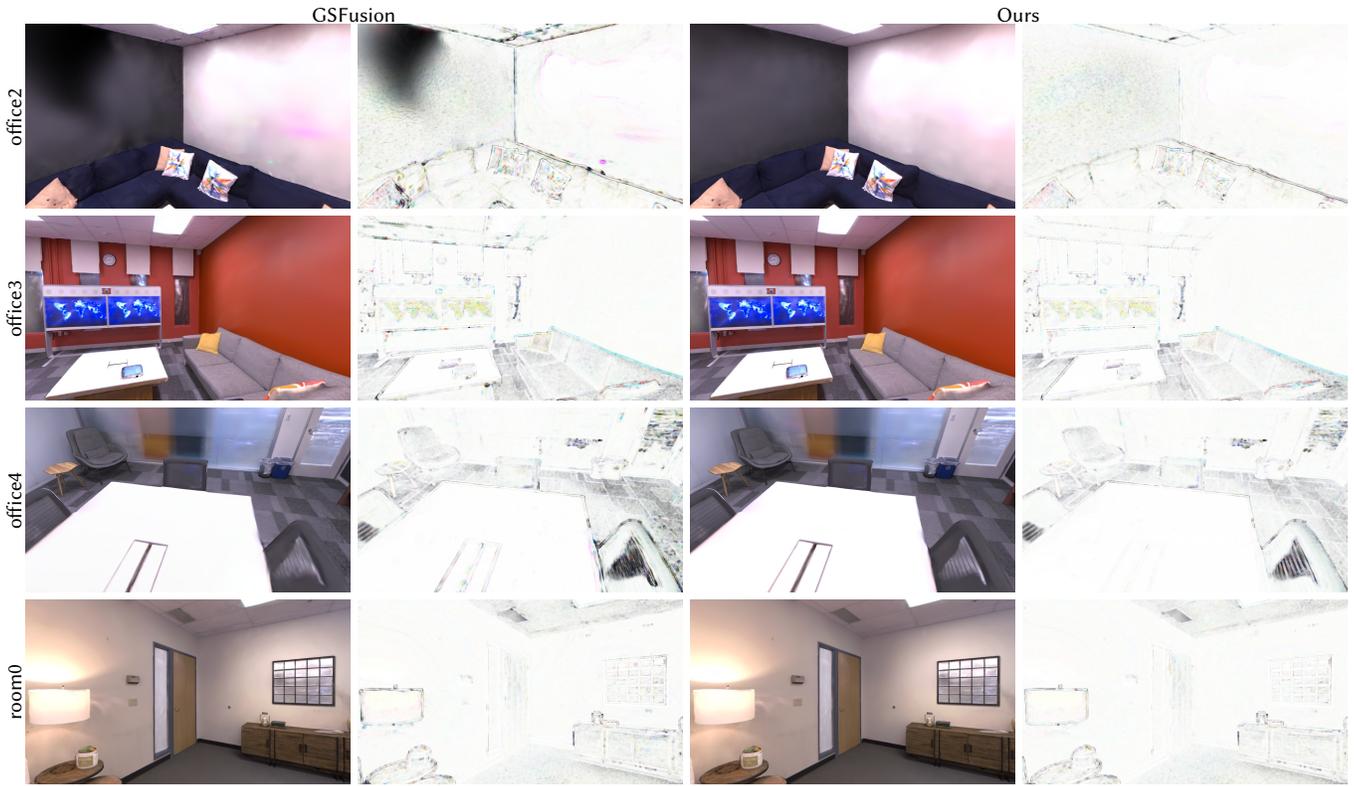

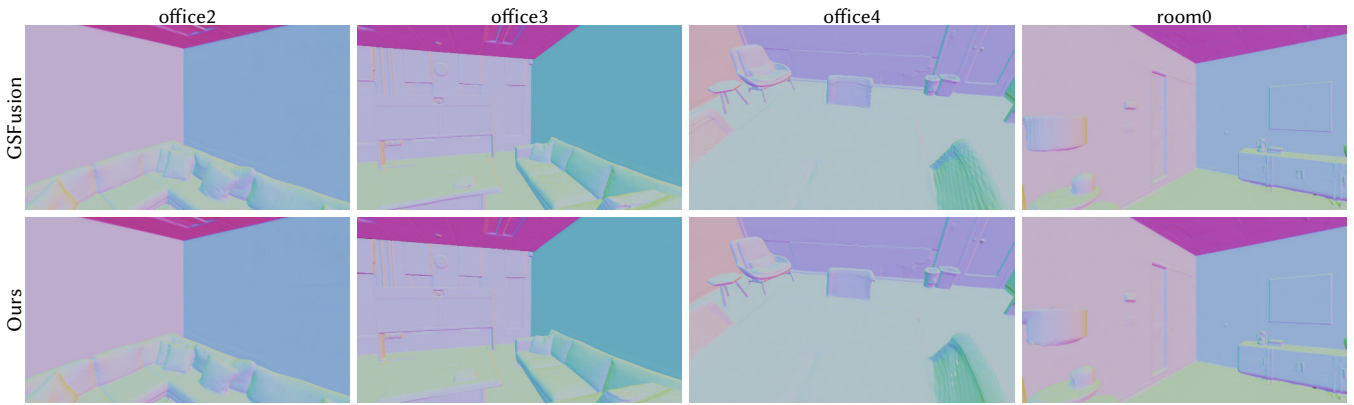
\begin{figure*}[htbp]
    \centering
    \small\sffamily
\begingroup%
\makeatletter%
\providecommand\color[2][]{%
	\errmessage{(Inkscape) Color is used for the text in Inkscape, but the package 'color.sty' is not loaded}%
	\renewcommand\color[2][]{}%
}%
\providecommand\transparent[1]{%
	\errmessage{(Inkscape) Transparency is used (non-zero) for the text in Inkscape, but the package 'transparent.sty' is not loaded}%
	\renewcommand\transparent[1]{}%
}%
\providecommand\rotatebox[2]{#2}%
\newcommand*\fsize{\dimexpr\f@size pt\relax}%
\newcommand*\lineheight[1]{\fontsize{\fsize}{#1\fsize}\selectfont}%
\ifx\svgwidth\undefined%
	\setlength{\unitlength}{506.92204285bp}%
	\ifx\svgscale\undefined%
		\relax%
	\else%
		\setlength{\unitlength}{\unitlength * \real{\svgscale}}%
	\fi%
\else%
	\setlength{\unitlength}{\svgwidth}%
\fi%
\global\let\svgwidth\undefined%
\global\let\svgscale\undefined%
\makeatother%
\begin{picture}(1,0.29718666)%
	\lineheight{1}%
	\setlength\tabcolsep{0pt}%
	\put(0,0){\includegraphics[width=\unitlength,page=1]{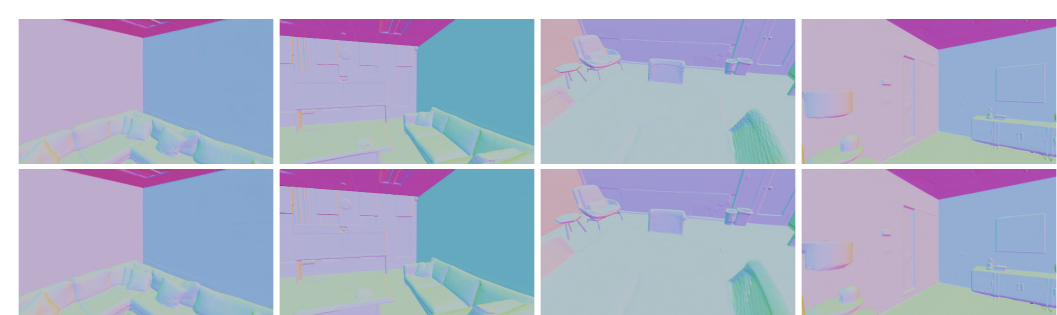}}%
	\put(0.01463734,0.21090596){\color[rgb]{0,0,0}\rotatebox{90}{\makebox(0,0)[t]{\lineheight{1.25}\smash{\begin{tabular}[t]{c}GSFusion\end{tabular}}}}}%
	\put(0.01463734,0.06832931){\color[rgb]{0,0,0}\rotatebox{90}{\makebox(0,0)[t]{\lineheight{1.25}\smash{\begin{tabular}[t]{c}Ours\end{tabular}}}}}%
	\put(0.13817748,0.28219429){\color[rgb]{0,0,0}\makebox(0,0)[t]{\lineheight{1.25}\smash{\begin{tabular}[t]{c}office2\end{tabular}}}}%
	\put(0.38525769,0.28219429){\color[rgb]{0,0,0}\makebox(0,0)[t]{\lineheight{1.25}\smash{\begin{tabular}[t]{c}office3\end{tabular}}}}%
	\put(0.63233808,0.28219429){\color[rgb]{0,0,0}\makebox(0,0)[t]{\lineheight{1.25}\smash{\begin{tabular}[t]{c}office4\end{tabular}}}}%
	\put(0.87941903,0.28219429){\color[rgb]{0,0,0}\makebox(0,0)[t]{\lineheight{1.25}\smash{\begin{tabular}[t]{c}room0\end{tabular}}}}%
\end{picture}%
\endgroup%

    \caption{\textbf{Qualitative reconstruction results on \rgbd~Datasets.}The images show the mapping results on office2, office3, office4, and room0 of Replica dataset \cite{straub2019replica}. We compare our method against GSFusion\cite{wei2024gsfusion}.}
    \label{fig:qualitative_recon_gsfusion}
    \Description{}
\end{figure*}

To evaluate the rendering performance of our pipeline, we integrate our voxel grid with a Gaussian Splatting (GS) renderer and compare against GSFusion \cite{wei2024gsfusion}, a recent method designed for efficient point-based rendering. \tabref{tab:render_gs} reports standard perceptual metrics, \ac{psnr}, \ac{ssim}, and \ac{lpips}, as well as rendering speed in \ac{fps}. Our multi-resolution configuration achieves consistently better results across all quality metrics. This improvement is attributed to our adaptive voxel grid, which maintains high-resolution detail in complex regions while avoiding redundant splat generation in homogeneous areas. This leads to more accurate reconstructions and better perceptual fidelity in rendered images. In terms of performance, our base version achieves the highest \ac{fps} (28.05), demonstrating the strength of our fully GPU-based quad-tree rendering mechanism. Unlike prior methods that rely on fixed splat densities or CPU-side pre-processing, our implementation leverages a parallel quad-tree structure to dynamically manage splat density at runtime. This enables both high-quality and real-time rendering, at the expense of efficiency and visual quality. In \tabref{tab:gsfusion_recon}, we provide a comparison of reconstruction results with GSFusion, noting that it uses Supereight2 \cite{funk2021multiresolution} as the underlying mapping pipeline. In \figref{fig:qualitative_rendering} and \figref{fig:qualitative_recon_gsfusion}, we present qualitative examples of NVS quality and 3D reconstruction quality, respectively. 
Both pipelines required approximately 5962 MB for the Gaussian representation. Additionally, our method used only 486 MB to store the map and quad-tree, accounting for just 8\% of the total GPU memory needed for the entire process, demonstrating efficient resource use despite the added adaptive structures.  For mapping alone, GSFusion takes 10.39 ms, while our approach takes 6.81 ms.

\section{Ablation on Multi-resolution}
Our approach employs a hash-based voxel grid in which spatial resolution is adaptively controlled by local \ac{tsdf} variance. This enables the system to allocate finer voxels in geometrically complex or detailed regions while using coarser voxels in smoother or less informative areas, leading to substantial reductions in memory usage.
To assess the effectiveness of this adaptive-resolution strategy, we compare our method with a single-resolution baseline in \tabref{tab:reconstruction_multi}. The baseline employs a uniform voxel size of 1 cm for \rgbd~data and 20 cm for \lidar~data. In contrast, our multi-resolution configuration adapts voxel sizes between 1 cm (finest) and 2 cm (coarsest) for \rgbd, and between 20 cm (finest) and 40 cm (coarsest) for \lidar. 
As expected, the single-resolution setup achieves slightly higher reconstruction accuracy due to uniformly fine sampling, serving as an upper bound. However, our multi-resolution variant achieves nearly equivalent reconstruction quality with significantly fewer mesh vertices and faces, translating into meaningful memory savings. This illustrates a key trade-off between geometric precision and memory efficiency, which is particularly advantageous in large-scale or resource-constrained environments.

Interestingly, in the rendering setup \tabref{tab:render_gs}, the multi-resolution configuration outperforms even the fixed-resolution baseline in perceptual quality, because the adaptive voxel structure spawns more splats in detailed regions and fewer in flat areas. Consequently, we achieve sharper renderings and improved metrics while maintaining high frame rates. These benefits are further enhanced by our GPU-accelerated quad-tree, which enables efficient real-time splat management.
Overall, these results confirm our variance-driven resolution control as a flexible and effective strategy for balancing accuracy and memory. The approach proves well-suited for a range of real-time applications, including dense 3D reconstruction and photorealistic rendering, where performance and resource constraints are critical.

\subsection{Variance Thresholds}
Since our adaptive resolution scheme depends directly on the local \ac{tsdf} variance, the choice of the variance threshold $\sigma$ is critical in balancing between reconstruction accuracy and memory usage. Lower thresholds favor the allocation of finer voxels over larger areas, improving geometric precision at the cost of increased memory. Conversely, higher thresholds limit fine-resolution allocation to regions of steep \ac{tsdf} variation, reducing resource usage but potentially omitting small-scale details. To investigate this trade-off, we evaluate different values of the variance threshold $\sigma$ on the Replica dataset \cite{straub2019replica}.
\figref{fig:ablation_threshold} reports their impacts on reconstruction quality metrics, while \figref{fig:res_comparison} qualitatively illustrates how varying the threshold changes the distribution of coarse and fine voxels in the scene. The results confirm the expectation: very low thresholds yield the most accurate reconstructions but allocate significantly more fine-resolution voxels, whereas moderate thresholds keep the reconstruction quality comparable, with reduced memory usage.
\section{Conclusion}
We presented a novel multi-resolution voxel grid framework for real-time 3D reconstruction that adapts resolution based on the local variance of \ac{tsdf} observations. Unlike traditional octree or VDB-based methods, our approach leverages a flat hash table to manage voxel blocks of varying resolution, enabling constant-time access and full GPU parallelization. This design supports efficient integration and rendering while remaining sensor-agnostic, handling both \rgbd~and \lidar~inputs. Our experiments demonstrate that the proposed method achieves competitive reconstruction quality, with up to 4× memory savings and 13× speedup compared to fixed-resolution baselines. The integration with a GPU-accelerated quad-tree further enables high-performance rendering through Gaussian Splatting, balancing visual fidelity and efficiency. By combining adaptive resolution, scalability, and real-time performance, our method offers a practical and extensible solution for large-scale 3D mapping in graphics and robotics applications. We release our implementation as open-source to encourage further research and adoption.
\begin{acks}
This work has been partially supported by PNRR MUR project PE0000013-FAIR
\end{acks}
\balance
\bibliographystyle{ACM-Reference-Format}
\bibliography{main}

@STRING{tro      = {IEEE Trans. on Robotics} }

@STRING{tog =  {ACM Trans. on Graphics} }

@STRING{arxiv   = {arXiv preprint} }

@STRING{cvpr    = {Proc.~of the IEEE Conf.~on Computer Vision and Pattern 
Recognition (CVPR)} }

@STRING{eccv    = {Proc.~of the Europ.~Conf.~on Computer Vision (ECCV)} }

@STRING{iccv    = {Proc.~of the IEEE Intl.~Conf.~on Computer Vision (ICCV)} }

@STRING{icra    = {Proc.~of the IEEE Intl.~Conf.~on Robotics \& Automation 
(ICRA)} }

@STRING{ijrr    = {Intl.~Journal~of Robotics Research (IJRR)} }

@STRING{iros    = {Proc.~of the IEEE/RSJ Intl.~Conf.~on Intelligent Robots and 
Systems (IROS)} }

@STRING{ismar   = {Proc.~of the Intl.~Symposium~on Mixed and Augmented Reality 
(ISMAR)} }

@STRING{ral     = {IEEE Robotics and Automation Letters (RA-L)} }

@STRING{rss     = {Proc.~of Robotics: Science and Systems (RSS)} }

@STRING{sensors = {IEEE Sensors Journal} }

@STRING{tro     = {IEEE Trans.~on Robotics (TRO)} }

@STRING{threedv = {Proc. of the International Conference on 3D Vision~(3DV)}}

@STRING{siggraph = {Proc.~of the ACM SIGGRAPH Conf.~on Computer Graphics and Interactive Techniques} }

@STRING{siggraphtalks = {ACM SIGGRAPH Talks} }

@STRING{eg = {Proc.~of Eurographics (EG)} }

@STRING{vmv = {Proc.~of the Vision, Modeling, and Visualization Workshop (VMV)} }

@STRING{tip = {IEEE Trans.~on Image Processing (TIP)} }

@STRING{jggt = {Journal of Graphics, GPU, and Game Tools (JGGGT)} }

@article{niessner2013real,
  title={Real-time 3D reconstruction at scale using voxel hashing},
  author={Nie{\ss}ner, Matthias and Zollh{\"o}fer, Michael and Izadi, Shahram and Stamminger, Marc},
  journal=tog,
  volume={32},
  number={6},
  pages={1--11},
  year={2013},
}

@article{museth2013vdb,
  title={VDB: High-resolution sparse volumes with dynamic topology},
  author={Museth, Ken},
  journal=tog,
  volume={32},
  number={3},
  pages={1--22},
  year={2013},
}

@inproceedings{curless1996volumetric,
author = {Curless, Brian and Levoy, Marc},
title = {A volumetric method for building complex models from range images},
year = {1996},
isbn = {0897917464},
url = {https://doi.org/10.1145/237170.237269},
doi = {10.1145/237170.237269},
booktitle = siggraph,
pages = {303–312},
numpages = {10},
}

@inproceedings{lorensen1987marching,
author = {Lorensen, William E. and Cline, Harvey E.},
title = {Marching cubes: A high resolution 3D surface construction algorithm},
year = {1987},
isbn = {0897912276},
url = {https://doi.org/10.1145/37401.37422},
doi = {10.1145/37401.37422},
booktitle = siggraph,
pages = {163–169},
numpages = {7},
}

@article{kerbl20233d,
  title={3d gaussian splatting for real-time radiance field rendering.},
  author={Kerbl, Bernhard and Kopanas, Georgios and Leimk{\"u}hler, Thomas and Drettakis, George},
  journal=tog,
  volume={42},
  number={4},
  pages={139--1},
  year={2023}
}

@inproceedings{oleynikova2017voxblox,
  title={Voxblox: Incremental 3d euclidean signed distance fields for on-board mav planning},
  author={Oleynikova, Helen and Taylor, Zachary and Fehr, Marius and Siegwart, Roland and Nieto, Juan},
  booktitle=iros,
  pages={1366--1373},
  year={2017},
  organization={IEEE}
}

@article{muller2022instant,
  title={Instant neural graphics primitives with a multiresolution hash encoding},
  author={M{\"u}ller, Thomas and Evans, Alex and Schied, Christoph and Keller, Alexander},
  journal=tog,
  volume={41},
  number={4},
  pages={1--15},
  year={2022},
}

@inproceedings{zheng2024mapadapt,
  title = {{{MAP-ADAPT}}: {{Real-Time Quality-Adaptive Semantic 3D Maps}}},
  shorttitle = {{{MAP-ADAPT}}},
  booktitle = eccv,
  author = {Zheng, Jianhao and Barath, Daniel and Pollefeys, Marc and Armeni, Iro},
  editor = {Leonardis, Ale{\v s} and Ricci, Elisa and Roth, Stefan and Russakovsky, Olga and Sattler, Torsten and Varol, G{\"u}l},
  year = {2024},
  pages = {220--237},
  doi = {10.1007/978-3-031-72933-1_13},
  isbn = {978-3-031-72933-1},
  langid = {english}
}

@inproceedings{vespa2019adaptive,
  title={Adaptive-resolution octree-based volumetric SLAM},
  author={Vespa, Emanuele and Funk, Nils and Kelly, Paul HJ and Leutenegger, Stefan},
  booktitle=threedv,
  pages={654--662},
  year={2019},
  organization={IEEE}
}

@article{vizzo2022vdbfusion,
  title = {{{VDBFusion}}: {{Flexible}} and {{Efficient TSDF Integration}} of {{Range Sensor Data}}},
  shorttitle = {{{VDBFusion}}},
  author = {Vizzo, Ignacio and Guadagnino, Tiziano and Behley, Jens and Stachniss, Cyrill},
  year = {2022},
  month = feb,
  journal = {Sensors},
  volume = {22},
  number = {3},
  pages = {1296},
  issn = {1424-8220},
  doi = {10.3390/s22031296},
  urldate = {2025-05-16},
}

@article{wei2024gsfusion,
  title = {{{GSFusion}}: {{Online RGB-D Mapping Where Gaussian Splatting Meets TSDF Fusion}}},
  shorttitle = {{{GSFusion}}},
  author = {Wei, Jiaxin and Leutenegger, Stefan},
  year = {2024},
  month = dec,
  journal =ral,
  volume = {9},
  number = {12},
  pages = {11865--11872},
  issn = {2377-3766},
  doi = {10.1109/LRA.2024.3502065},
  urldate = {2025-02-18},
}

@inproceedings{amanatides1987fast,
  title = {A {{Fast Voxel Traversal Algorithm}} for {{Ray Tracing}}},
  author = {Amanatides, John and Woo, Andrew},
  year = {1987},
  issn = {1017-4656},
  urldate = {2025-05-20},
  langid = {english},
  booktitle = eg,
  DOI = {10.2312/egtp.19871000}
}

@inproceedings{teschner2003optimized,
  title = {Optimized {{Spatial Hashing}} for {{Collision Detection}} of {{Deformable Objects}}},
  booktitle = vmv,
  author = {Teschner, Matthias and Heidelberger, Bruno and M{\"u}ller, Matthias and Pomerantes, Danat and Gross, Markus H.},
  editor = {Ertl, Thomas},
  year = {2003},
  pages = {47--54},
}

@article{funk2021multiresolution,
  title = {Multi-{{Resolution 3D Mapping With Explicit Free Space Representation}} for {{Fast}} and {{Accurate Mobile Robot Motion Planning}}},
  author = {Funk, Nils and Tarrio, Juan and Papatheodorou, Sotiris and Popovi{\'c}, Marija and Alcantarilla, Pablo F. and Leutenegger, Stefan},
  year = {2021},
  month = apr,
  journal = ral,
  volume = {6},
  number = {2},
  pages = {3553--3560},
  issn = {2377-3766},
  doi = {10.1109/LRA.2021.3061989},
  urldate = {2025-05-20}
}

@article{welford1962note,
  title = {Note on a {{Method}} for {{Calculating Corrected Sums}} of {{Squares}} and {{Products}}},
  author = {Welford, B. P.},
  year = {1962},
  month = aug,
  journal = {Technometrics},
  volume = {4},
  number = {3},
  pages = {419--420},
  issn = {0040-1706},
  doi = {10.1080/00401706.1962.10490022},
  urldate = {2025-05-22}
}

@article{vespa2018efficient,
  title = {Efficient {{Octree-Based Volumetric SLAM Supporting Signed-Distance}} and {{Occupancy Mapping}}},
  author = {Vespa, Emanuele and Nikolov, Nikolay and Grimm, Marius and Nardi, Luigi and Kelly, Paul H. J. and Leutenegger, Stefan},
  year = {2018},
  month = apr,
  journal = ral,
  volume = {3},
  number = {2},
  pages = {1144--1151},
  issn = {2377-3766},
  doi = {10.1109/LRA.2018.2792537},
  urldate = {2025-02-18}
}

@inproceedings{newcombe2011kinectfusion,
  title = {{{KinectFusion}}: {{Real-time}} Dense Surface Mapping and Tracking},
  shorttitle = {{{KinectFusion}}},
  booktitle = ismar,
  author = {Newcombe, Richard A. and Izadi, Shahram and Hilliges, Otmar and Molyneaux, David and Kim, David and Davison, Andrew J. and Kohi, Pushmeet and Shotton, Jamie and Hodges, Steve and Fitzgibbon, Andrew},
  year = {2011},
  month = oct,
  pages = {127--136},
  doi = {10.1109/ISMAR.2011.6092378},
  urldate = {2025-05-23}
}

@article{pan2024pinslam,
  title = {{{PIN-SLAM}}: {{LiDAR SLAM Using}} a {{Point-Based Implicit Neural Representation}} for {{Achieving Global Map Consistency}}},
  shorttitle = {{{PIN-SLAM}}},
  author = {Pan, Yue and Zhong, Xingguang and Wiesmann, Louis and Posewsky, Thorbj{\"o}rn and Behley, Jens and Stachniss, Cyrill},
  year = {2024},
  journal = tro,
  volume = {40},
  pages = {4045--4064},
  issn = {1941-0468},
  doi = {10.1109/TRO.2024.3422055},
  urldate = {2025-03-06}
}

@article{song2024n3mapping,
  title = {N{\textsuperscript{3}}-{{Mapping}}: {{Normal Guided Neural Non-Projective Signed Distance Fields}} for {{Large-Scale 3D Mapping}}},
  shorttitle = {N{\textsuperscript{3}}-{{Mapping}}},
  author = {Song, Shuangfu and Zhao, Junqiao and Huang, Kai and Lin, Jiaye and Ye, Chen and Feng, Tiantian},
  year = {2024},
  month = jun,
  journal = ral,
  volume = {9},
  number = {6},
  pages = {5935--5942},
  issn = {2377-3766},
  doi = {10.1109/LRA.2024.3396638},
  urldate = {2024-10-07}
}

@inproceedings{huang2023neural,
  title = {Neural {{Kernel Surface Reconstruction}}},
  booktitle = cvpr,
  author = {Huang, Jiahui and Gojcic, Zan and Atzmon, Matan and Litany, Or and Fidler, Sanja and Williams, Francis},
  year = {2023},
  month = jun,
  pages = {4369--4379},
  doi = {10.1109/CVPR52729.2023.00425},
  urldate = {2025-03-05},
  copyright = {https://doi.org/10.15223/policy-029},
  isbn = {979-8-3503-0129-8},
  langid = {english}
}

@inproceedings{zhu2024nicerslam,
  title = {{{NICER-SLAM}}: {{Neural Implicit Scene Encoding}} for {{RGB SLAM}}},
  shorttitle = {{{NICER-SLAM}}},
  booktitle = threedv,
  author = {Zhu, Zihan and Peng, Songyou and Larsson, Viktor and Cui, Zhaopeng and Oswald, Martin R. and Geiger, Andreas and Pollefeys, Marc},
  year = {2024},
  month = mar,
  pages = {42--52},
  issn = {2475-7888},
  doi = {10.1109/3DV62453.2024.00096},
  urldate = {2025-05-23}
}

@inproceedings{museth2021nanovdb,
  title = {{{NanoVDB}}: {{A GPU-Friendly}} and {{Portable VDB Data Structure For Real-Time Rendering And Simulation}}},
  shorttitle = {{{NanoVDB}}},
  booktitle = siggraphtalks,
  author = {Museth, Ken},
  year = {2021},
  month = aug,
  pages = {1--2},
  doi = {10.1145/3450623.3464653},
  urldate = {2025-05-23},
}

@article{zhang2021multicamera,
  title = {Multi-{{Camera LiDAR Inertial Extension}} to the {{Newer College Dataset}}},
  author = {Zhang, Lintong and Camurri, Marco and Fallon, M.},
  year = {2021},
  month = dec,
  journal = {ArXiv},
  urldate = {2025-03-06},
}

@article{tao2025oxford,
  title = {The {{Oxford Spires Dataset}}: {{Benchmarking}} Large-Scale {{LiDAR-visual}} Localisation, Reconstruction and Radiance Field Methods},
  shorttitle = {The {{Oxford Spires Dataset}}},
  author = {Tao, Yifu and {Mu{\~n}oz-Ba{\~n}{\'o}n}, Miguel {\'A}ngel and Zhang, Lintong and Wang, Jiahao and Fu, Lanke Frank Tarimo and Fallon, Maurice},
  year = 2025,
  month = sep,
  journal = ijrr,
  pages = {02783649251369905},
  issn = {0278-3649},
  doi = {10.1177/02783649251369905},
  langid = {english},
}

@inproceedings{dai2017scannet,
  title = {{{ScanNet}}: {{Richly-Annotated 3D Reconstructions}} of {{Indoor Scenes}}},
  shorttitle = {{{ScanNet}}},
  booktitle = cvpr,
  author = {Dai, Angela and Chang, Angel X. and Savva, Manolis and Halber, Maciej and Funkhouser, Thomas and Nie{\ss}ner, Matthias},
  year = {2017},
  month = jul,
  pages = {2432--2443},
  issn = {1063-6919},
  doi = {10.1109/CVPR.2017.261},
  urldate = {2025-05-13},
}

@misc{straub2019replica,
  title = {The {{Replica Dataset}}: {{A Digital Replica}} of {{Indoor Spaces}}},
  shorttitle = {The {{Replica Dataset}}},
  author = {Straub, Julian and Whelan, Thomas and Ma, Lingni and Chen, Yufan and Wijmans, Erik and Green, Simon and Engel, Jakob J. and {Mur-Artal}, Raul and Ren, Carl and Verma, Shobhit and Clarkson, Anton and Yan, Mingfei and Budge, Brian and Yan, Yajie and Pan, Xiaqing and Yon, June and Zou, Yuyang and Leon, Kimberly and Carter, Nigel and Briales, Jesus and Gillingham, Tyler and Mueggler, Elias and Pesqueira, Luis and Savva, Manolis and Batra, Dhruv and Strasdat, Hauke M. and Nardi, Renzo De and Goesele, Michael and Lovegrove, Steven and Newcombe, Richard},
  year = {2019},
  month = jun,
  number = {arXiv:1906.05797},
  eprint = {1906.05797},
  primaryclass = {cs},
  doi = {10.48550/arXiv.1906.05797},
  urldate = {2025-05-20},
}

@inproceedings{mescheder2019occupancy,
  title = {Occupancy {{Networks}}: {{Learning 3D Reconstruction}} in {{Function Space}}},
  shorttitle = {Occupancy {{Networks}}},
  booktitle = cvpr,
  author = {Mescheder, Lars and Oechsle, Michael and Niemeyer, Michael and Nowozin, Sebastian and Geiger, Andreas},
  year = {2019},
  month = jun,
  pages = {4455--4465},
  issn = {2575-7075},
  doi = {10.1109/CVPR.2019.00459},
  urldate = {2025-05-23},
}

@article{lengyel2010transition,
  title = {Transition {{Cells}} for {{Dynamic Multiresolution Marching Cubes}}},
  author = {Lengyel, Eric},
  year = {2010},
  month = may,
  journal = jggt,
  volume = {15},
  number = {2},
  pages = {99--122},
  issn = {2151-237X, 2151-2272},
  doi = {10.1080/2151237X.2011.563682},
  urldate = {2025-05-24},
  langid = {english}
}

@article{williams2024fvdb,
  title = {{{fVDB}} : {{A Deep-Learning Framework}} for {{Sparse}}, {{Large Scale}}, and {{High Performance Spatial Intelligence}}},
  shorttitle = {{{fVDB}}},
  author = {Williams, Francis and Huang, Jiahui and Swartz, Jonathan and Klar, Gergely and Thakkar, Vijay and Cong, Matthew and Ren, Xuanchi and Li, Ruilong and {Fuji-Tsang}, Clement and Fidler, Sanja and Sifakis, Eftychios and Museth, Ken},
  year = {2024},
  month = jul,
  journal = tog,
  volume = {43},
  number = {4},
  pages = {133:1--133:15},
  issn = {0730-0301},
  doi = {10.1145/3658226},
}

@article{wang2004image,
  title = {Image Quality Assessment: From Error Visibility to Structural Similarity},
  shorttitle = {Image Quality Assessment},
  author = {Wang, Zhou and Bovik, A.C. and Sheikh, H.R. and Simoncelli, E.P.},
  year = {2004},
  month = apr,
  journal = tip,
  volume = {13},
  number = {4},
  pages = {600--612},
  issn = {1941-0042},
  doi = {10.1109/TIP.2003.819861}
}

@inproceedings{zhang2018unreasonable,
  title = {The {{Unreasonable Effectiveness}} of {{Deep Features}} as a {{Perceptual Metric}}},
  booktitle = cvpr,
  author = {Zhang, Richard and Isola, Phillip and Efros, Alexei A. and Shechtman, Eli and Wang, Oliver},
  year = {2018},
  pages = {586--595},
}

@article{dai2017bundlefusion,
  title = {{{BundleFusion}}: {{Real-Time Globally Consistent 3D Reconstruction Using On-the-Fly Surface Reintegration}}},
  shorttitle = {{{BundleFusion}}},
  author = {Dai, Angela and Nie{\ss}ner, Matthias and Zollh{\"o}fer, Michael and Izadi, Shahram and Theobalt, Christian},
  year = {2017},
  month = jun,
  journal = tog,
  volume = {36},
  number = {3},
  pages = {1--18},
  issn = {0730-0301, 1557-7368},
  doi = {10.1145/3054739},
  langid = {english},
}

@inproceedings{keetha2024splatam,
  title = {{{SplaTAM}}: {{Splat}}, {{Track}} \& {{Map 3D Gaussians}} for {{Dense RGB-D SLAM}}},
  shorttitle = {{{SplaTAM}}},
  booktitle = cvpr,
  author = {Keetha, Nikhil and Karhade, Jay and Jatavallabhula, Krishna Murthy and Yang, Gengshan and Scherer, Sebastian and Ramanan, Deva and Luiten, Jonathon},
  year = {2024},
  month = jun,
  pages = {21357--21366},
  doi = {10.1109/CVPR52733.2024.02018},
  copyright = {https://doi.org/10.15223/policy-029},
  isbn = {979-8-3503-5300-6},
  langid = {english},
}

@inproceedings{matsuki2024gaussian,
  title = {Gaussian {{Splatting SLAM}}},
  booktitle = cvpr,
  author = {Matsuki, Hidenobu and Murai, Riku and Kelly, Paul H. J. and Davison, Andrew J.},
  year = {2024},
  month = jun,
  pages = {18039--18048},
  issn = {2575-7075},
  doi = {10.1109/CVPR52733.2024.01708},
}

@inproceedings{peng2024rtgslam,
  title = {{{RTG-SLAM}}: {{Real-time 3D Reconstruction}} at {{Scale}} Using {{Gaussian Splatting}}},
  shorttitle = {{{RTG-SLAM}}},
  booktitle = siggraph,
  author = {Peng, Zhexi and Shao, Tianjia and Liu, Yong and Zhou, Jingke and Yang, Yin and Wang, Jingdong and Zhou, Kun},
  year = {2024},
  month = jul,
  pages = {1--11},
  doi = {10.1145/3641519.3657455},
  isbn = {979-8-4007-0525-0},
}

@inproceedings{brizi2024vbr,
  title = {{{VBR}}: {{A Vision Benchmark}} in {{Rome}}},
  shorttitle = {{{VBR}}},
  booktitle = icra,
  author = {Brizi, Leonardo and Giacomini, Emanuele and Giammarino, Luca Di and Ferrari, Simone and Salem, Omar and De Rebotti, Lorenzo and Grisetti, Giorgio},
  year = 2024,
  month = may,
  pages = {15868--15874},
  doi = {10.1109/ICRA57147.2024.10611395},
}

@article{digiammarino2023photometric,
  title = {Photometric {{LiDAR}} and {{RGB-D Bundle Adjustment}}},
  author = {Di Giammarino, Luca and Giacomini, Emanuele and Brizi, Leonardo and Salem, Omar and Grisetti, Giorgio},
  year = 2023,
  month = jul,
  journal = ral,
  volume = {8},
  number = {7},
  pages = {4362--4369},
  issn = {2377-3766},
  doi = {10.1109/LRA.2023.3281907},
}

@inproceedings{giacomini2025splatloam,
  title = {Splat-{{LOAM}}: {{Gaussian Splatting LiDAR Odometry}} and {{Mapping}}},
  shorttitle = {Splat-{{LOAM}}},
  booktitle = iccv,
  author = {Giacomini, Emanuele and Di Giammarino, Luca and De Rebotti, Lorenzo and Grisetti, Giorgio and Oswald, Martin R.},
  year = 2025,
  pages = {27630--27639},
  langid = {english},
}

@article{ferrari2024madicp,
  title = {{{MAD-ICP}}: {{It}} Is {{All About Matching Data}} -- {{Robust}} and {{Informed LiDAR Odometry}}},
  shorttitle = {{{MAD-ICP}}},
  author = {Ferrari, Simone and Giammarino, Luca Di and Brizi, Leonardo and Grisetti, Giorgio},
  year = 2024,
  month = nov,
  journal = ral,
  volume = {9},
  number = {11},
  pages = {9175--9182},
  issn = {2377-3766},
  doi = {10.1109/LRA.2024.3456509},
}

@inproceedings{zhang2014loam,
  title = {{{LOAM}}: {{Lidar Odometry}} and {{Mapping}} in {{Real-time}}},
  shorttitle = {{{LOAM}}},
  booktitle = rss,
  author = {Zhang, Ji and Singh, Sanjiv},
  year = 2014,
  month = jul,
  doi = {10.15607/RSS.2014.X.007},
  isbn = {978-0-9923747-0-9},
  langid = {english},
}

@inproceedings{schops2019bad,
  title = {{{BAD SLAM}}: {{Bundle Adjusted Direct RGB-D SLAM}}},
  shorttitle = {{{BAD SLAM}}},
  booktitle = cvpr,
  author = {Sch{\"o}ps, Thomas and Sattler, Torsten and Pollefeys, Marc},
  year = 2019,
  month = jun,
  pages = {134--144},
  issn = {2575-7075},
  doi = {10.1109/CVPR.2019.00022},
}

@article{cwian2025madba,
  title = {{{MAD-BA}}: {{3D LiDAR Bundle Adjustment}} -- {{From Uncertainty Modelling}} to {{Structure Optimization}}},
  shorttitle = {{{MAD-BA}}},
  author = {{\'C}wian, Krzysztof and Giammarino, Luca Di and Ferrari, Simone and Ciarfuglia, Thomas and Grisetti, Giorgio and Skrzypczy{\'n}ski, Piotr},
  year = 2025,
  month = jul,
  journal = ral,
  volume = {10},
  number = {7},
  pages = {7254--7261},
  issn = {2377-3766},
  doi = {10.1109/LRA.2025.3573628}
}

@inproceedings{digiammarino2022mdslam,
  title = {{{MD-SLAM}}: {{Multi-cue Direct SLAM}}},
  shorttitle = {{{MD-SLAM}}},
  booktitle = iros,
  author = {Di Giammarino, Luca and Brizi, Leonardo and Guadagnino, Tiziano and Stachniss, Cyrill and Grisetti, Giorgio},
  year = 2022,
  month = oct,
  pages = {11047--11054},
  issn = {2153-0866},
  doi = {10.1109/IROS47612.2022.9981147},
}

\clearpage
\end{document}